\documentclass[numberedappendix]{emulateapj}
\usepackage{amssymb,amsmath}
\def\bs{\boldsymbol}

\def\gsim{\;\rlap{\lower 2.5pt
\hbox{$\sim$}}\raise 1.5pt\hbox{$>$}\;}
\def\lsim{\;\rlap{\lower 2.5pt
\hbox{$\sim$}}\raise 1.5pt\hbox{$<$}\;}
\makeatletter
\newcommand{\vast}{\bBigg@{3}}
\newcommand{\Vast}{\bBigg@{5}}
\makeatother

\newcommand{\Rmnum}[1]{\expandafter\@slowromancap\romannumeral #1@}

\begin{document}
\title{Turbulence-Induced Relative Velocity of Dust particles IV:  the Collision Kernel}

\author{Liubin Pan\altaffilmark{1} \& Paolo Padoan\altaffilmark{2}}
\altaffiltext{1}{Harvard-Smithsonian Center for Astrophysics,
60 Garden St., Cambridge, MA 02138; {\tt lpan@cfa.harvard.edu}}
\altaffiltext{2}{ICREA \& Institut de Ci\`{e}ncies del Cosmos, Universitat de Barcelona, IEEC-UB, Mart\'{i} Franqu\`{e}s 1, 
E08028 Barcelona, Spain; {\tt ppadoan@icc.ub.edu}}

\begin{abstract}

Motivated by its importance for modeling dust particle growth in protoplanetary disks, we study 
turbulence-induced collision statistics of inertial particles as a function of the particle friction time, 
$\tau_{\rm p}$. We show that turbulent clustering significantly enhances the collision rate for 
particles of similar sizes with $\tau_{\rm p}$ corresponding to the inertial range of the flow. 
If the friction time, $\tau_{\rm p,h}$, of the larger particle is in the inertial range, the collision kernel per 
unit cross section increases with increasing friction time, $\tau_{\rm p,l}$, of the smaller particle, and 
reaches the maximum at $\tau_{\rm p,l} = \tau_{\rm p,h}$, where the clustering effect peaks. 
This feature is not captured by commonly-used kernel formula, which neglects the effect of clustering. 
We argue that turbulent clustering helps alleviate the bouncing barrier problem for planetesimal 
formation. We also investigate the collision velocity statistics using a collision-rate weighting 
factor to account for higher collision frequency for particle pairs with larger relative velocity. 
For $\tau_{\rm p,h}$ in the inertial range, the rms relative velocity with collision-rate weighting 
is found to be invariant with $\tau_{\rm p,l}$ and scales with $\tau_{\rm p,h}$ roughly as 
$\propto \tau_{\rm p,h}^{1/2}$. The weighting factor favors collisions with larger relative velocity, 
and including it leads to more destructive and less sticking collisions. We compare two collision 
kernel formulations based on spherical and cylindrical geometries. The two formulations give 
consistent results for the collision rate and the collision-rate weighted statistics, 
except that the spherical formulation predicts more head-on collisions than the cylindrical formulation. 

\end{abstract}

\section{Introduction}

Modeling the growth of dust particles in turbulent protoplanetary disks is of crucial importance for 
understanding the challenging problem of planetesimal formation. While the classical picture for 
planetesimal formation by gravitational instability in a dense dust layer at the disk mid plane suffers 
from self-generated turbulence (see Chiang \& Youdin 2010 for a review), the mechanism via the 
collisional growth of dust grains from (sub)micron size  to kilometer size is frustrated by the meter-size barrier. 
As the particles grow to decimeter sizes, they become less sticky and, depending on the particle 
properties and the collision velocity, the collisions may result in bouncing or fragmentation 
(Blum \& Wurm 2008; G\"uttler et al. 2010). This would suppress the further growth of the 
particles past the meter size.  
Meanwhile, the meter-size particles have large radial drift velocities and may rapidly 
get lost to the central star on a timescale of a hundred years (Weidenschilling 1977). 
The viability of planetesimal formation by dust coagulation depends on how fast and to 
which size dust particles can grow by collisions. 

To answer this question, an accurate evaluation of the collision statistics is required. In a series of 
recent papers (Pan \& Padoan 2013, Pan et al.\ 2014a, 2014b), we have investigated the statistics 
of the relative velocity of dust particles induced by turbulent motions. In Pan \& Padoan (2013) 
and Pan et al.\  (2014a), we conducted a systematic study of the root-mean-square (rms) of 
turbulence-induced relative velocity using both numerical and analytical approaches. 
In particular, we showed that  the prediction of the Pan \& Padoan (2010) model for the rms relative velocity 
is in good agreement with the simulation data, confirming the physical picture of the model.  
In Pan et al.\ (2014b), we analyzed the probability distribution of the relative velocity as a function 
of the particle friction time, and discussed  its important role in determining the fractions of collisions 
leading to sticking, bouncing and fragmentation (Windmark et al.\ 2012, Garaud et al.\ 2013). 
  
In this fourth paper, we focus on the turbulence-induced collision kernel, extending the earlier 
result on equal-size particles, known as the monodisperse case (Pan \& Padoan 2013, hereafter 
Paper I), to the general bidisperse case for different particles of arbitrary sizes. 
The collision kernel is needed to estimate the collision rate of dust particles. 
Coagulation models for dust particles in protoplanetary disks set the kernel to be the product 
of the cross section and the root-mean-square (rms) relative velocity of the colliding particles. These studies commonly 
adopt  the model of V\"olk et al.\ (1980) and its later developments for the rms collision velocity 
(e.g., Markiewicz et al.\ 1991, Cuzzi \& Hogan 2003, Ormel \& Cuzzi 2007). 
For convenience, we will refer to this type of models generally as the V\"olk-type model. 

There are several uncertainties in the commonly-used prescriptions for dust 
particle collisions. First, the effect of turbulent clustering (Maxey 1987) 
on the collision kernel was typically not accounted for. 
Second, most theoretical models, including V\"olk et al.\ (1980) and our model (Pan \& Padoan 2010), 
only predict the 2nd order moment (rms) of the relative velocity, while the calculations of 
the collision kernel and the average collision energy require the 1st and the 3rd oder moments respectively 
(e.g., Hubbard 2012, 2013). Finally,  the accuracy of the V\"olk-type model and the kernel formulation 
have not been systematically tested against numerical simulations. An assessment 
of the V\"olk-type model is of particular interest, considering a weakness of its physical picture raised 
in our earlier works\footnote{For example, the model of V\"olk-type  does not keep track of the 
separation of the two particles backward in time, which we argued is of crucial importance in 
determining the correlation in the velocities of the two particles induced by turbulent eddies of a given size (Pan \& Padoan 2010, 2013).} (Pan \& Padoan 2010, 2013). 
In this paper, we will focus on evaluating the collision kernel and the 
collision velocity statistics using a numerical simulation and on clarifying the 
first two uncertainties in the commonly-used kernel formulation. We defer a thorough assessment 
of the accuracy of V\"olk-type models for 
the rms relative velocity to a later work.

In \S 2, we describe the simulation used in this work. \S 3 presents  numerical results for the collision kernel, including 
the effects of turbulent clustering and turbulent-induced relative velocity. 
In \S 4, we analyze the collision velocity statistics weighted by the collision rate, which accounts 
for the higher collision frequency for particle pairs with larger relative velocity.  
\S 5 discusses the implication of our numerical results  on dust particle growth in protoplanetary disks. 
Our conclusions are summarized in \S 6. 

\section{Numerical Simulation}

We analyze the simulation data of Pan \& Padoan (2013; hereafter Paper I).  
In the simulation, we evolved a weakly compressible flow on a 
uniform 512$^3$ periodic grid and integrated 
the trajectories of inertial particles in a wide size range using the Pencil code
(Brandenburg \& Dobler 2002, Johansen et al.\ 2004). 
The turbulent flow was driven and maintained by a large-scale solenoidal force generated 
in Fourier space using all modes with wave length larger than half box size. 
The three components of each mode are independently drawn from Gaussian distributions of 
equal variance. The direction for each mode is random, and the driving force is 
statistically refreshed at each time step. This conventional driving 
scheme produces a turbulent flow with the broadest inertial range 
possible at a given resolution and the maximum degree of statistical isotropy. 
At steady state, the rms Mach number, $M$, of the simulated 
flow is $\simeq 0.1$, consistent with turbulence conditions in protoplanetary 
disks. There has been compelling evidence that the behavior of structure functions of all orders in 
a subsonic turbulent flow with $M \lsim 1$ is almost identical to incompressible turbulence 
(e.g., Porter et al.\ 2002, Padoan et al.\ 2004, Pan \& Scannapieco 2011). This suggests that, at $M\lsim 1$, the 
particle collision statistics would be insensitive to the Mach number, and our results are applicable 
for any subsonic turbulent flow.  

The regular and Taylor Reynolds numbers of the simulated flow are $\simeq 1000$ 
and $\simeq 200$, respectively. The 1D rms  flow velocity is $u' \simeq 6.8 u_\eta$, 
with $u_\eta$ the Kolmogorov velocity. The integral scale, $L$, of the flow is about 1/6 
of the box size, and the Kolmogorov scale is $\eta \simeq 0.6$ grid cell size. 
The large eddy time, $T_{\rm eddy} \equiv L/u'$, was estimated to be $20 \tau_\eta$, 
where $\tau_\eta$ is the Kolmogorov timescale, corresponding to the smallest eddies in the flow. 
Using tracer particles, we also measured  the Lagrangian correlation time, $T_{\rm L} \simeq 14.4\tau_\eta$. 
As shown in Paper I, $T_{\rm L}$ is more relevant than $T_{\rm eddy}$ for 
understanding the particle velocity. We refer the reader to 
our earlier papers (Papers I \& II) for a detailed description of 
the statistics of the simulated flow velocity field, ${\bs u}$. 

In the simulated flow, we evolved 14 particle species of different sizes, 
each containing 33.6 million particles. The particle inertia is characterized by 
the friction or stopping time, $\tau_{\rm p}$, which ranges from $0.1 \tau_{\eta}$ to 
$43\, T_{\rm L}$. 
Defining the Stokes number as $St\equiv \tau_{\rm p}/\tau_\eta$, 
this range corresponds to $ 0.1 \le St \le800$. We compute 
the relative velocity, ${\bs w} \equiv {\bs v}^{(1)} -  {\bs v}^{(2)}$, 
of nearby particles (1) and (2) (with velocities ${\bs v}^{(1,2)}$) at a small separation, 
${\bs r}$, and analyze the collision statistics as a function of 
the Stokes number pairs ($St_1$, $St_2$).  
We denote the Stokes numbers of the smaller and larger particles as $St_{\ell}$ and $St_{h}$, respectively. 
It is convenient to show the collision kernel and velocity 
as functions of the friction time, $\tau_{\rm ph}$ (or the Stokes number $St_{h}$) of the 
larger particle, and the friction time ratio $f$, defined as $f\equiv St_{\ell}/St_{h}$. By definition, $0\le f\le 1$. 
The method used to analyze the particle statistics was 
given  in Paper I, to which we refer the reader for details.  

An interesting question concerning our simulation is whether the particle 
collision statistics depends on the driving scheme adopted for the turbulent flow. 
Considering the universality of turbulent structures at small scales, 
we expect that the collisions of relatively small particles with $\tau_{\rm p}$ 
much below $T_{\rm eddy}$ (or $T_{\rm L}$) are not affected by 
the driving mechanism. The same is expected for the large particle limit 
with $\tau_{\rm p} \gg  T_{\rm eddy}$. The motions of these large particles 
are similar to Brownian motion because even the largest eddies in the flow 
would act like random kicks, when viewed on the friction time, $\tau_{\rm p}$, of these 
particles. Therefore, collisions of particles in the $\tau_{\rm p} \gg  T_{\rm eddy}$ limit 
are also insensitive to the details of the 
velocity structures at the driving scale.
However, for intermediate particles with $\tau_{\rm p}$ close to $T_{\rm eddy}$, the dynamics is coupled 
to the flow velocity structures around the driving scale, which are non-universal, and 
thus their collision statistics may depend on how the flow is driven.
An exploration of this possible dependence requires a number of numerical 
simulations using different driving schemes, which are computationally expensive and 
beyond the scope of the current work.  

\section{The Collision Kernel}

The collision rate per unit volume between two particle species with Stokes numbers
$St_1$ and $St_2$ can be expressed as $\bar{n}_1 \bar{n}_2 \Gamma$,
where $\bar{n}_1$ and $\bar{n}_2$ are the average number densities and $\Gamma$ 
is the collision kernel (see Zhou et al.\ 2001). In Saffman \& Turner (1956), 
two formulations, named the spherical and cylindrical formulations, were proposed for 
$\Gamma$ (see also Wang et al.\ 2000; Paper I). In the spherical formulation, the kernel depends on the radial component, 
$w_{\rm r} = {\bs w} \cdot {\bs r}/r$, of the relative velocity, ${\bs w}$, of two particles 
at a separation of ${\bs r}$. In practical applications, the kernel should be evaluated 
at $r$ equal to the sum, $d$($\equiv a_{\rm p1} + a_{\rm p2} $), of the radii ($a_{\rm p1,2}$) of the two particles. 
The spherical kernel assumes that, in a time interval of $dt$, 
the number of collisions for a given particle (1) with particles (2) approaching at a radial velocity of $w_{\rm r}$ 
is determined by the number of particles (2) in a spherical shell of radius  $d$ and thickness 
$w_{\rm r} dt$ around particle (1). In this picture, the kernel $\Gamma^{\rm sph}=  -4 \pi  d^2 g( d, St_1, St_2)  \int_{-\infty}^0  w_{\rm r}P (w_{\rm r}) d w_{\rm r}$, 
where $g$ is the radial distribution function (RDF) and  $P$ is the probability distribution function (PDF)  of $w_{\rm r}$. 

The RDF, $g$, reflects the effect of particle clustering, and is defined such that the average number of particles (2) 
in a volume $dV$ at a distance, $r$, from a reference particle (1) is $\bar{n}_2 g(r) dV$. If the particles are  uniformly distributed, we have 
$g=1$. Only particle pairs approaching each other ($w_{\rm r} \le 0$) are counted in the spherical 
formulation. From statistical stationarity, $-\int_{-\infty}^0  w_{\rm r}P (w_{\rm r}) d w_{\rm r} = \int_{0}^{\infty}  w_{\rm r}P (w_{\rm r}) d w_{\rm r}$, 
meaning that the numbers of particles (2) moving toward and away from a particle (1) are equal on average.  
Therefore, the spherical kernel can be rewritten as $\Gamma^{\rm sph}= 2 \pi  d^2 g(d) \langle |w_{\rm r}|\rangle$, 
where $\langle |w_{\rm r}|\rangle$ ($\equiv \int_{-\infty}^{\infty}   |w_{\rm r}|  P (w_{\rm r}) d w_{\rm r}$) 
(Wang et al.\ 2000).   For dust particles, $d$ is much smaller than the Kolmogorov scale 
of protoplanetary turbulence, thus well beyond the reach of simulations. 
We will measure the kernel at $r$ much larger than the actual particle size 
and examine its convergence with decreasing $r$.

The cylindrical formulation assumes that the number of collisions a particle (1) encounters in a time interval $dt$ is 
equal to the number of particles (2) in a cylinder of length $\langle |{\bs w}|\rangle dt$ and radius $d$, 
where $\langle |{\bs w}|\rangle $ the average of the 3D amplitude, $|{\bs w}|$, of the 
relative velocity. The formulation gives $\Gamma^{\rm cyl} =  \pi  d^2 g(d) \langle |{\bs w} |\rangle$. Here  $\langle | {\bs w} |\rangle$
is the average of the 3D amplitude, $|{\bs w} |$, of the relative velocity,  which can be calculated
from the PDF, $P(|{\bs w}|)$, of $|{\bs w}|$ as $\langle | {\bs w} |\rangle\equiv \int_{0}^{\infty}  |{\bs w}| 
P (|{\bs w}|) d |{\bs  w}|$. 
The cylindrical formulation does not distinguish approaching 
and separating particle pairs, and the spherical formulation 
appears to be physically more reasonable. Wang et al.\ (1998) argued that 
the spherical formulation is valid in general. 
In kinetic theory, the velocities of colliding molecules are completely 
random and independent, and the two formulations give exactly 
the same estimates for the collision rate (Wang et al.\ 1998). This is, 
however, not true in general. 

The PDFs, $P (w_{\rm r})$ and $P(|{\bs w}|)$, can be expressed in 
terms of the distribution, $P_{\rm v} ({\bs w}) $, of the relative velocity 
vector, ${\bs w}$. $P(|{\bs w}|)$ is related to $P_{\rm v} ({\bs w})$ 
as $P(|{\bs w}|) = \int_{0}^{\pi}  \sin(\theta)d\theta  \int_{0}^{2\pi}  d\phi |{\bs w}|^2 
P_{\rm v} (|{\bs w}|, \theta, \phi)$, where $\theta$ is the angle between ${\bs w}$ 
and the particle separation ${\bs r}$. For the radial velocity PDF, we have $P (w_{\rm r}) = \int_{-\infty}^{\infty} d{w_{\rm t1}} 
\int_{-\infty}^{\infty}  d{w_{\rm t2}}  P_{\rm v} ({w_{\rm r}},  {w_{\rm t1}}, {w_{\rm t2}})$,
where $w_{\rm t1}$ and $w_{\rm t2}$ are the two tangential components of ${\bs w}$.

If the distribution of the direction of ${\bs w}$ is isotropic with respect to 
the particle separation, ${\bs r}$, $P_{\rm v} ({\bs w})$ 
is a function of $|{\bs w}|$ only, i.e., $ P_{\rm v}(|{\bs w}|, \theta, \phi)= P_{\rm v} (|{\bs w}|) = P_{\rm v} ((w_{\rm r}^2 + w_{\rm t1}^2+ w_{\rm t2}^2 )^{1/2}) $.
In that case, $P(|{\bs w}|) = 4\pi  |{\bs w}|^2  P_{\rm v}(|{\bs w}|)$, and it is easy to show that, 
\begin{equation}
P (w_{\rm r})=2\pi\int_{|w_{\rm r}|}^{\infty}  |{\bs w}| P_{\rm v}(|{\bs w}|) d|{\bs w}|.
\label{wrpdf}
\end{equation}
Integrating $|w_{\rm r}| P(w_{\rm r})$ using eq.\ (\ref{wrpdf}), we find that
$\langle |w_{\rm r}|\rangle = 2\pi \int_{0}^{\infty} |{\bs w}|^3 P_{\rm v}(|{\bs w}|) d|{\bs w}|$.
Therefore, we have $\langle |{\bs w} |\rangle = 2\langle |w_{\rm r}|\rangle$, meaning 
that $\Gamma^{\rm cyl}$ and $\Gamma^{\rm sph}$ are equal. 
Wang et al.\ (1998) proved that the spherical and cylindrical formulations give 
equivalent results for the collision rate if the three components of the relative velocity are uncorrelated 
and Gaussian with equal variance. However, these strong assumptions are not necessary. 
Our proof above only requires the isotropy of the direction of ${\bs w}$. 

The isotropy of ${\bs w}$ is observed for most particle species in our simulation 
expect for small particles of similar size with Stokes numbers $ St \lsim 1$ (see Papers I \& II). 
For these small particles, the rms of a tangential component of ${\bs w}$ perpendicular to ${\bs r}$ 
is larger than the radial component. The physical origin of this anisotropy 
is that in incompressible turbulence the transverse structure function 
of the velocity field is larger than the longitudinal one (see, e.g., Monin \& Yaglom 1975). 
For example, for small particles of equal size, the relative velocity 
depends on the local flow velocity difference, ${\Delta {\bs u}}$, 
across the particle distance, and thus inherits the inequality of the 
radial and tangental components of  ${\Delta {\bs u}}$ (see Papers I \& II). 
Due to the anisotropy of the direction of ${\bs w}$, the spherical and 
cylindrical kernels are not equal for particle of similar sizes with $St \lsim 1$. 
As discussed in Papers I \& II, the isotropy of  ${\bs w}$ improves with increasing $St$, decreasing $f$, and decreasing 
$r$. A detailed explanation for these trends can be found in  Papers I \& II. 

Wang et al.\ (2000) found  that, for particles of equal-size with $St \lsim 1$, the cylindrical kernel is larger than 
the spherical one by $\lsim 20\%$\footnote{In Paper I, we computed both $\Gamma^{\rm sph}$ and $\Gamma^{\rm cyl}$ 
for particles of equal sizes and showed that they almost coincide with a difference $\lsim 5\%$ at $St \lsim 1$.}. 
For larger particles, the two formulations give the same collision rate, as the direction of ${\bs w}$ 
becomes isotropic. By comparing the two formulations to the directly measured collision rate, Wang et al.\ (2000) 
showed that the spherical kernel is more accurate for $St \lsim 1$ particles.
In our data analysis, we will consider both the spherical and cylindrical formulations, 
and compare the collision statistics computed from the two formulations. 
We will also compare the two formulations with the kernel formula commonly used in dust coagulation models.  

\subsection{The Radial Distribution Function}

Inertial particles suspended in turbulent flows show inhomogeneous 
spatial distribution. A well-known explanation for the inhomogeneous distribution 
is that inertial particles tend to be expelled from vortical structures 
in the flow (Maxey 1987; Squires \& Eaton 1991).  
Vortices induce rotation of the particles, leading to a centrifugal force 
that pushes the particles out.  Particles are thus collected in regions 
in between strong vortices, where the strain of the flow velocity dominates over the vorticity 
(see Fig.\ 2 of Cencini et al.\ 2006 and Fig.\ 3 of Pan et al.\ 2011). 

\begin{figure}[t]
\centerline{\includegraphics[height=3in]{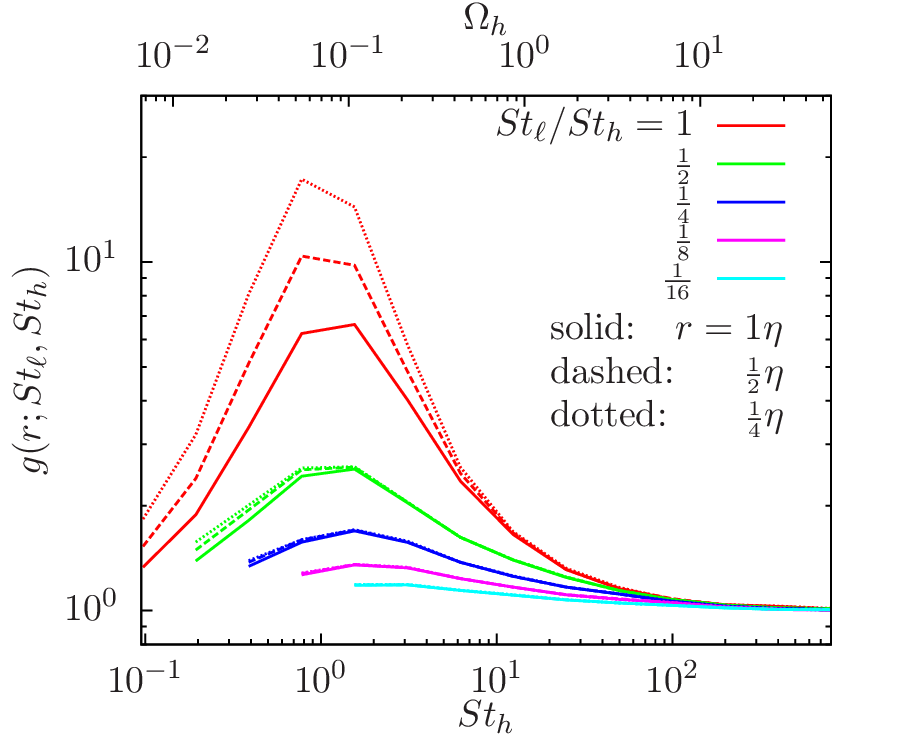}}
\caption{The RDF as a function of the larger Stokes number, $St_{h}$, and the Stokes ratio, $f$.
Lines of different colors show different values of $f$ 
in the range  $\frac{1}{16} \le f \le 1$. 
 }
\label{rdf} 
\end{figure}

The degree of clustering can be quantified by the RDF, $g$. In Fig.\ \ref{rdf}, we plot the 
RDFs at fixed Stokes ratios, $f$, as a function of  the Stokes number, $St_{h}$, of the larger particle 
at three distances, $r$. The upper X-axis normalizes the friction 
time, $\tau_{\rm p,h}$, of the larger particle to $T_{\rm L}$, i.e., $\Omega_{h} = \tau_{\rm p,h}/T_{\rm L}$.
The red lines correspond to the case of equal-size particles, 
which has been already shown in Paper I. 
The monodisperse RDF peaks for $St\simeq1$ particles, 
whose friction time couples to the smallest eddies of the flow.  
In Pan et al.\ (2011), we showed that the peak of the RDF at $St\simeq1$ can be explained by 
an analysis of the effective compressibility of the particle ``flow", which is the largest at $St\simeq 1$. 
The RDF of equal-size particles with $St\lsim6$ shows a significant 
dependence on $r$, increasing as a power law with decreasing $r$ (see Pan et al.\ 2011 and Paper I). 
For these particles, the relative velocity decreases with decreasing $r$ 
(see Fig.\ \ref{absv} in \S 3.2). 
The slower relative motions  at smaller scales make the spatial dispersion (or diffusion) 
of the particle concentration less efficient, leading to an increase of the RDF toward 
smaller $r$. The $r-$dependence disappears at $St \gsim 6$, where the particle relative 
velocity becomes $r-$independent (see Fig.\ \ref{absv}).

At a given $St_{h}$, the RDF decreases with decreasing Stokes ratio $f$. 
In general, the RDF between two different particles is smaller than the monodisperse 
RDF of each particle, i.e., $g(r, St_{\ell}, St_{h}) \le g(r, St_{\ell}, St_{\ell})$ 
and $g(r, St_{\ell}, St_{h}) \le g(r, St_{h}, St_{h})$ (Zhou et al.\ 2001).  
This is because particles of different sizes tend to cluster at different 
locations. If one continuously changes the particle size, 
the positions of maximum clustering intensity would shift spatially (see Pan et al.\ 2011). 
As the size difference of the two particles increases, 
the spatial separation between their local concentration peaks 
becomes larger, leading to a decreases in their ``relative" clustering. 

The RDF of different particles has also a much weaker $r$-dependence 
than for particles of the same size. As shown in Pan et al.\ (2011), the 
bidisperse RDF as a function of $r$ becomes flat and approaches a
constant at small $r$ (see also Chun et al.\ 2005). 
Intuitively, the increases of the RDF toward small $r$ would 
eventually stop at scales below the typical separation between local concentration 
peaks of two different particle species\footnote{This can also be understood from the 
theoretical model of Chun et al.\ (2005).  Different responses of particles of different 
sizes to the flow velocity provides a contribution, known as the acceleration contribution (see Saffman \& Turner 1956, Wang et al.\ 2000 and 
Papers I \& II),  to their relative velocity. This $r-$independent contribution to the particle relative motions 
causes an extra spatial relative diffusion of the two particle species, 
which tends to reduce the clustering intensity and flatten the RDF toward small $r$.}. 
The flattering of the RDF toward small $r$ makes it easier to achieve convergence 
for particles of different sizes. At $f\le1/2$, the RDF already converges at 
$r\gsim \eta/4$, except for the smallest two particles ($St_{\ell}=0.1$ and $St_{h}=0.19$) in our simulation.

To our knowledge, the effect of turbulent clustering on the collision 
kernel has not been included in existing dust coagulation models.  As will be 
discussed in \S 3.4 and \S 5,  neglecting this effect tends to underestimate the collision rate.

\subsection{The Radial Relative Velocity}

\begin{figure}[t]
\center{\includegraphics[height=2.9in]{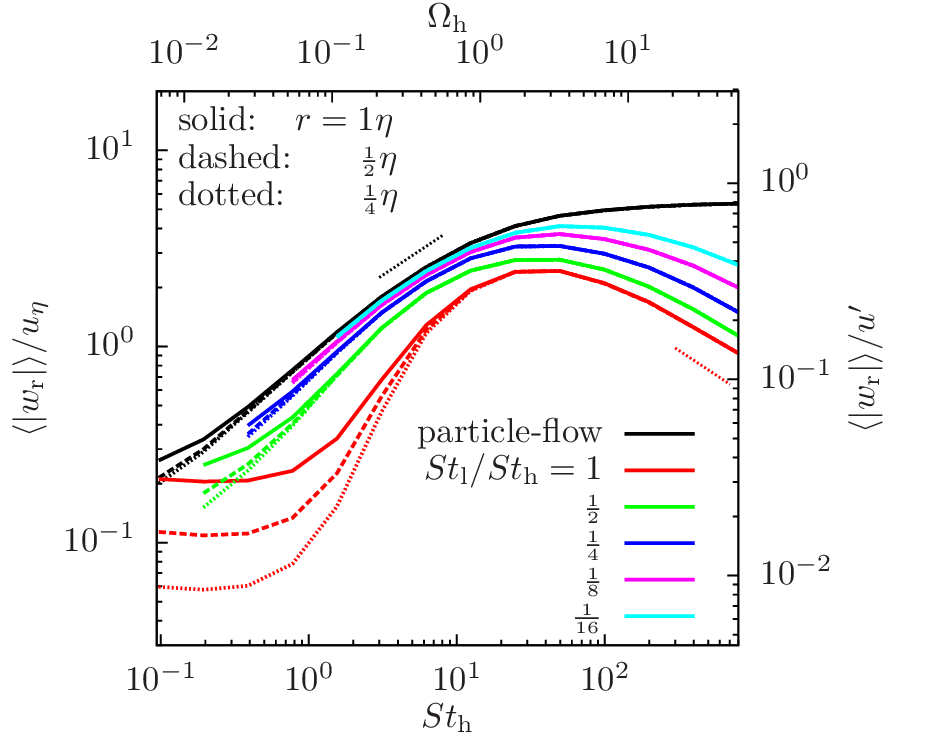}}
\caption{Absolute average, $\langle |w_{\rm r}| \rangle$, of the radial relative 
speed normalized to $u_\eta$ (left Y-axis) and $u'$ (right Y-axis).
Black lines are the particle-flow relative velocity, corresponding to $f=0$. 
Dotted line segments denote $St_{h}^{1/2}$ (black) and $St_{h}^{-1/2}$ (red) scalings.}
\label{absv} 
\end{figure}

In Fig.\ \ref{absv}, we show $\langle |w_{\rm r}| \rangle$ at fixed values of $f$. 
The red lines, corresponding to the monodisperse case, have been studied in Paper I 
(see Fig.\ 16 of Paper I), to which we refer the reader for details. The black lines plot the 
radial relative velocity, $\langle |w_{\rm f,r}| \rangle$, between 
the particle and the flow (or tracer) velocity at given distances,  corresponding 
to $St_{\ell}=0$ or $f=0$. 
All the curves for  $0<f<1$ lie in a region between the monodisperse 
and the particle-flow lines, which serve as useful delimiters.  

Most theoretical models for the particle relative velocity predict its variance 
$\langle w^2 \rangle$ (or $\langle w_{\rm r}^2 \rangle$ for the radial component), 
which is easier to model than the mean relative velocity, $\langle |\bs{w}|\rangle$ (or $\langle |w_{\rm r}| \rangle$). 
The variance (or the rms)  does not enter the collision kernel, 
but it is of theoretical importance, because understanding it helps reveal 
the underlying physics.    
The qualitative behavior of the mean radial relative velocity in Fig.\ \ref{absv} is very similar to the rms relative velocity 
shown in Figure 7 of Paper II for $\langle w^2 \rangle^{1/2}$ 
(see also the right panel of Fig.\ 10 in Paper II for the radial rms $\langle w_{\rm r}^2 \rangle^{1/2}$).
Therefore, the successful physical explanation for the 
rms relative velocity in Paper II based on the Pan \& Padoan model 
can also be used to understand the behavior of $\langle |w_{\rm r}| \rangle$ 
as a function of $St_h$ and $f$.

In the PP10 model, the relative velocity between two particles of 
any arbitrary sizes has two contributions, named the generalized shear and acceleration 
contributions\footnote{The terminology originates from Saffman and Turner (1956), who 
predicted that the radial relative velocity variance 
$\langle w_{\rm r}^2 \rangle = \frac{1}{15}(r/\tau_{\eta})^2 + [a (\tau_{\rm p,h} -\tau_{\rm p,l})]^2$ 
for small particles with $St_{\ell}, St_{h} \ll 1$, where $a$ is the rms acceleration 
of the flow. The two terms were referred to as the shear and acceleration terms. We thus 
named the two terms in the Pan \& Padoan model the generalized shear and 
acceleration terms, as they reduce to the two terms by Saffman and Turner (1956) 
in the small particle limit.  The contribution of the flow acceleration on the relative velocity of small particles of different 
sizes was also found by Weidenschilling (1984).  
}. 
The generalized shear contribution represents the effect of the 
particles' memory of the {\it spatial} flow velocity difference  ``seen" by 
the two particles at given times in the past. As the flow 
velocity difference, $\Delta u (\ell)$, scales with the length scale, 
$\ell$, the shear contribution depends on the separation, 
$d (t)$, of the two particles backward in time (i.e., at $t<0$). 
The memory timescale of an inertial particle is essentially its 
friction time, and the particles  ``forget"  the flow velocity they 
saw before a friction time ago, i.e., at $t\lsim -\tau_{\rm p}$.  Thus, considering 
the particle separation (and hence $\Delta u (d(t))$) 
increases backward in time, the flow velocity difference, $\Delta u$, across the 
particle separation, $d(-\tau_{\rm p})$, at a friction time ago plays a crucial role in 
determining the shear contribution.

The generalized acceleration term reflects different responses of different particles 
to the flow velocity and is associated with the {\it temporal} flow velocity 
difference, $\Delta_{\rm T} u$, along individual particle trajectories (see Fig.\ 1 of 
Paper II for an illustration). The generalized acceleration contribution increases with the friction time difference, $\tau_{\rm p, h}- \tau_{\rm p, l}$. 
For particles of equal size, it vanishes, and only the generalized shear term contributes  (Paper II).  
The generalized acceleration term is $r$-independent as it only depends on the temporal flow velocity statistics 
along {\it individual} particle trajectories (see Paper II for details). In Paper II, we showed that the prediction 
of the PP10 model for the rms relative velocity is in good agreement with the simulation results, 
confirming the physical picture of the model.

For particles of equal size, $\langle |w_{\rm r}| \rangle$ shows a strong $r-$
dependence at $St \lsim 1$. Due to the short memory/friction time, the particle distance $d(-\tau_{\rm p})$ 
a friction time ago is close to $r$, and thus $\Delta u (d(-\tau_{\rm p})) \simeq \Delta u(r)$, 
which is linear with $r$ for a small $r$. This gives rise to the $r$-dependence of the shear 
contribution. As $St$ increases, $d(-\tau_{\rm p})$ increases and becomes less sensitive 
to $r (= d(0))$. Thus, with increasing $St$, the relative velocity increases, and the $r-$dependence is 
weaker. At $St \gsim 6$, $\langle |w_{\rm r}| \rangle$ becomes $r-$ independent.  
Finally, for $\tau_{\rm p} \gg T_{\rm L}$, the flow velocity correlation/memory 
time is shorter than the particle memory, and the flow structures seen 
by the two particles within a friction time are incoherent.  
The action of the flow velocity on the particles is similar 
to random kicks in Brownian motion. This causes a reduction of the relative 
velocity by $(T_{\rm L}/\tau_{\rm p})^{1/2}$, leading to a $St^{-1/2}$ decrease in the $\tau_{\rm p} \gg T_{\rm L}$ limit.  
Intuitively, in this limit,  the motions of larger particles are slower, and their 
relative velocity decreases with increasing $\tau_{\rm p}$.   

At a given $St_{h}$,  $\langle |w_{\rm r}| \rangle$ increases with decreasing $f$. This is due to the 
generalized acceleration contribution, which increases
with increasing Stokes number difference. The PP10 model shows that the acceleration 
contribution dominates over the shear effect if the Stokes numbers differ by a factor of $\gsim4$ (Paper II). 
In the case of $f=0$, $w_{\rm f,r}$ can be estimated by the temporal flow velocity difference, 
$\Delta u_{\rm T} (\Delta \tau)$, at a time lag of $\Delta \tau \simeq \tau_{\rm p}$ along the particle trajectory 
(see Paper II).  Assuming that  $\Delta u_{\rm T}$ can be approximated by Lagrangian temporal velocity difference $\Delta u_{\rm L}$,
 and that  $\Delta u_{\rm L} \propto \Delta \tau^{1/2}$ for inertial-range time lags, this explains why $\langle |w_{\rm f,r}| \rangle$ 
 increases approximately as $St^{1/2}$ (dotted line segment) for $\tau_{\rm p, h}$ 
 in the inertial range. As $\Delta u_{\rm T}$ saturates at large time lags, $w_{\rm f,r}$ becomes
 constant for $\tau_{\rm p} \gg T_{\rm L}$. The $r-$dependence for particles of different sizes 
 is much weaker than the case of equal-size particles, as the acceleration contribution is $r-$independent. 
At $f \le 1/2$, $\langle |w_{\rm r}| \rangle$ already converges at $r=\frac{1}{4}\eta$. Together with the 
convergence of the RDF, this makes it easier to estimate the bidisperse collision kernel at $r\to 0$ 
than in the case of equal-size particles (see \S 3.3).  

\subsection{The Measured Collision Kernel}

\begin{figure*}[t]
\includegraphics[height=2.9in]{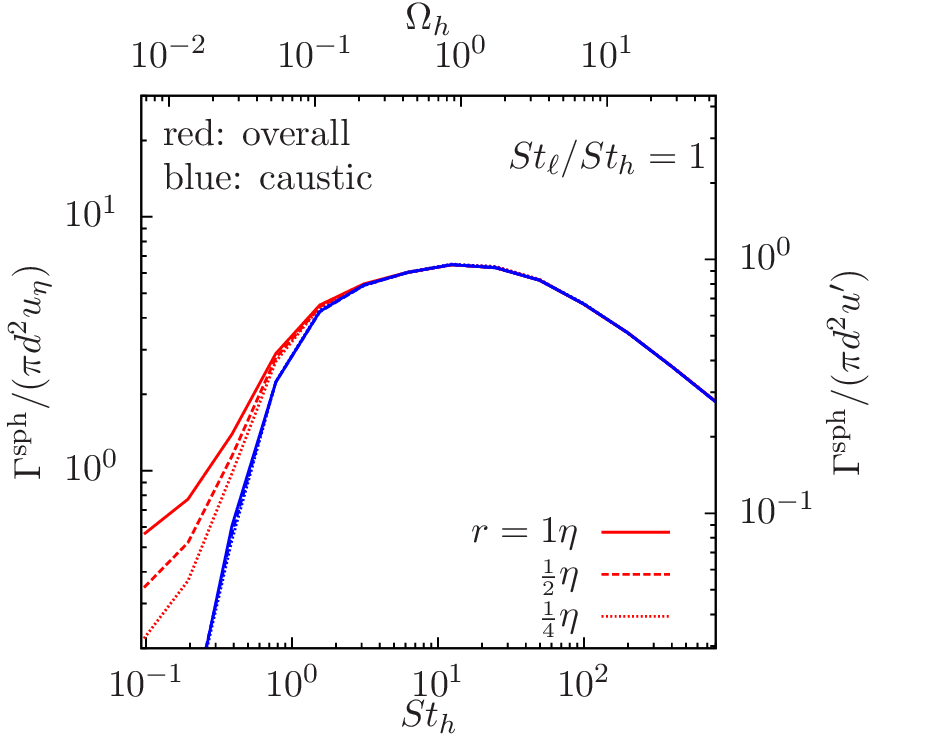}
\includegraphics[height=2.9in]{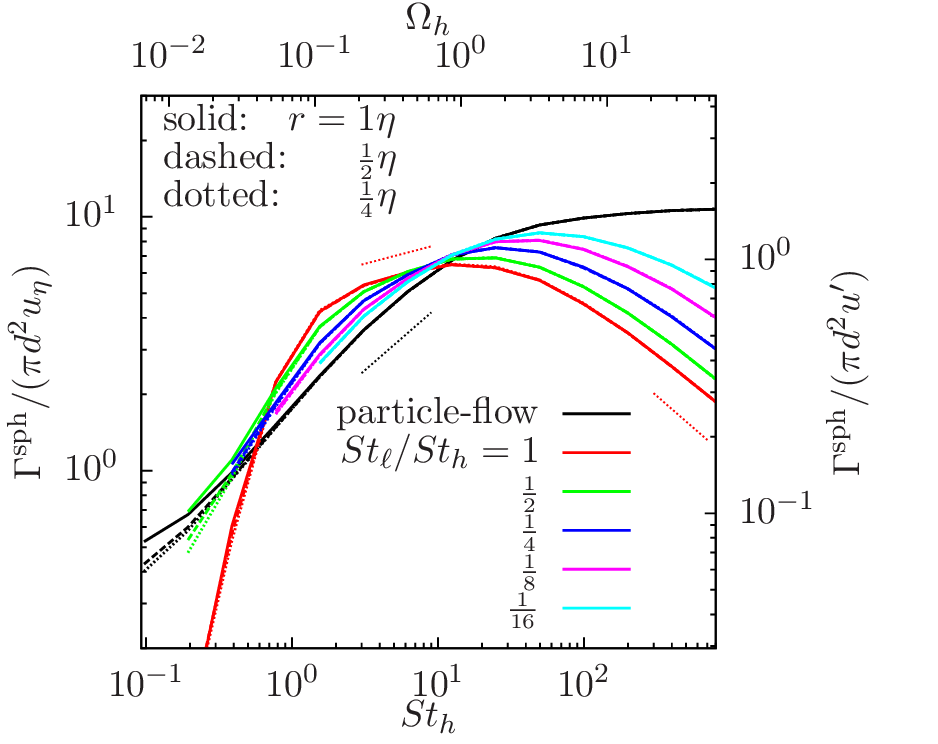}
\caption{Collision kernel per unit cross section in the spherical formulation. 
Left panel, the kernel for particles of equal sizes.  The red lines show the overall kernel measured 
at different $f$.  The blue lines are the $r$-independent contribution to the monodisperse kernel 
by caustic particle pairs with $w_{\rm r} \le -0.2 \,r/\tau_{\rm p}$. Right panel: the kernel for different 
Stokes ratios. Black lines plot the kernel between inertial and tracer particles ($f=0$). 
The red lines are the caustic contribution to the kernel for $f=1$, 
corresponding to the blue lines in the left panel.  The red dotted line 
segments denote $St^{0.15}$ (left) and $St^{-1/2}$ (right) scalings, and the black 
one shows a $St^{1/2}$ scaling.}
\label{kernel} 
\end{figure*}

In Fig.\  \ref{kernel}, we plot  the spherical collision kernel per unit cross section, 
$\Gamma^{\rm sph}/(\pi d^2) = 2 g \langle |w_{\rm r}| \rangle$, normalized to $u_{\eta}$ and $u'$ 
on the left  and right Y-axes, respectively. The left panel shows the kernel for equal-size particles with $f=1$. 
The red lines in this panel is essentially the product of the red lines in 
Fig.\ \ref{rdf} for $g$ and those in Fig.\ \ref{absv} for $\langle |w_{\rm r}|\rangle$.  
They correspond to the solid lines in Fig.\ 17 of Paper I. For convenience, we will also 
refer to the kernel per unit cross section as the normalized kernel. 

For  particles of equal size with $St\gsim 1$, the normalized 
kernel already converges at $r\gsim \frac{1}{4}\eta$, as the 
$r-$dependences of $g$ and $\langle |w_{\rm r}| \rangle$ cancel out  (see Paper I).  
However, for small particles with $St\lsim 1$, the kernel shows a significant $r-$dependence at $r \ge \frac{1}{4}\eta$. 
The decrease of $\langle |w_{\rm r}| \rangle$ with decreasing $r$ is faster 
than the increase of $g$, and the normalized kernel keeps decreasing in the $r$ 
range shown here. Since dust particle size in protoplanetary disks is much smaller than the 
Kolmogorov scale, the measured kernel for $St\lsim 1$ at $r\ge \frac{1}{4}\eta$ cannot be 
directly used in applications. One solution to this 
is to seek convergence of the normalized kernel by conducting larger 
simulations with a larger number of particles that allow accurate measurements 
at smaller scales.  Alternatively, one may try to extrapolate the measured kernel to 
the $r\to 0$ limit using the physical picture and theoretical models for the problem.
As the former is computationally expensive, we made an attempt in Paper I to pursue the latter approach.

Theoretical models have shown that, at a given small distances, $r$, 
there are two types of particle pairs, whose contributions to the collision kernel have different scaling 
behaviors with $r$ (see Falkovich et al.\ 2002, Wilkinson et al.\ 2006). 
Following the terminology of Wilkinson et al.\ (2006), 
we named the two types of particle pairs  ``continuous" and ``caustic" pairs in Paper I. 
As explained in details in Paper I, the two types correspond to the inner and outer parts of 
the probability distribution of $w_{\rm r}$ at small and large relative velocities, respectively.  

Physically, the continuous particle pairs are located in flow regions with small velocity gradients, 
where the particles can efficiently respond to the flow. Their relative velocity for $St \lsim1$ particles 
thus roughly follows the flow velocity difference across the particle distance. 
On the other hand, the caustic pairs correspond to flow regions with velocity gradients larger 
than $\simeq 1/\tau_{\rm p}$. In these regions, particle motions cannot catch up with the rapid 
flow velocity change, and their trajectories would deviate considerably from the flow elements. 
For example, particles may be shot out of the flow streamlines of high curvature (e.g., around a vortex), 
and these particles may cross the trajectories of nearby particles (Falkovich \& Pumir 2007). 
At the trajectory crossing point, the particle velocity becomes multivalued, giving rise to the formation of folds, 
named caustics, in the momentum-position phase space (see, e.g., Fig.\ 1 of Gustavsson \& Mehlig 2011). 
In contrast, the continuos pairs appear to be smooth in phase space. 
This difference in the phase space motivated the terminology for the two 
types of particle pairs (Wilkinson et al.\ 2006). 

The relative velocity of continuous particle pairs decreases linearly with 
decreasing $r$, faster than the increase of the RDF.  As a result, 
their contribution to the normalized kernel vanishes as $r \to 0$ 
(Gustavsson \& Mehlig 2011, Hubbard 2012; see Fig.\ 19 in Paper I). 
On the other hand, the caustic contribution, corresponding to trajectory crossing events, 
is predicted to be $r-$independent, and it becomes dominant at $r\to0$ (Gustavsson \& Melig 2011). 
We thus attempted to isolate the $r-$independent caustic contribution.

Motivated by previous theoretical studies and our simulation data, 
we use a critical radial relative velocity, $w_{\rm c}$($<0$), to split approaching 
particle pairs into continuous ($w_{\rm c} \le w_{\rm r} \le 0$) and caustic ($w_{\rm r} \le w_{\rm c}$) pairs.  
The contribution of caustic pairs to the kernel is then given by 
$\Gamma^{\rm cau} = -4\pi d^2 g \int_{-\infty}^{w_{\rm c}} w_{\rm r} P(w_{\rm r}) dw_{\rm r}$.  
Theoretical models suggest $w_{\rm c} \propto - r/\tau_{\rm p}$ (e.g., Falkovich et al.\ 2002; 
see PP10), but leave the coefficient in this estimate unfixed. There is thus a freedom for the 
choice of $w_{\rm c}$.  As the caustic contribution is expected to be $r-$independent, 
we aim to select a value of $w_{\rm c}$ that minimizes the $r-$dependence of  $\Gamma^{\rm cau}$. 
The best choice is found to be $w_{\rm c} = -0.2\, r/\tau_{\rm p}$. 
The blue lines in the left panel of Fig.\ \ref{kernel} plot $\Gamma^{\rm cau}$ with this $w_{\rm c}$, 
and we see that the three lines for different $r$ almost all collapse onto a single curve. 
In Paper I, we set $w_{\rm c} =- r/\tau_{\rm p}$, with which $\Gamma^{\rm cau}$ still had 
a (weak) $r-$dependence (see Fig.\  19 in Paper I). Setting $w_{\rm c} = -0.2\, r/\tau_{\rm p}$
appears to be a better choice for the purpose of obtaining  an $r-$independent 
kernel contribution.

In the right panel of Fig.\ \ref{kernel}, the red lines are the caustic contribution for  
particles of equal sizes, corresponding to the blue lines in the left panel. All the 
other lines in the right panel show the normalized kernel for particles of different sizes. 
Unlike the case of equal-size particles, the kernels for particles of different sizes with 
$f \le 1/2$ already converge at $r=\frac{1}{4}\eta$, and can thus be directly used in practical applications. 
As seen earlier, for $f\le \frac{1}{2}$, both the RDF and $\langle |w_{\rm r}| \rangle$ 
converge at $r=\frac{1}{4}\eta$. 

The black lines in the right panel plot the kernel between 
inertial particles and tracers, correspond to $f=0$. 
In this case,  the RDF is unity, and the normalized 
kernel is simply calculated as $2\langle |w_{\rm f,r}| \rangle$ with $\langle |w_{\rm f,r}| \rangle$ 
the average of the radial component of the particle-flow  relative velocity. 
The black lines and the red ones for $f=1$ provide useful limits, 
between which the curves for $0<f<1$ reside. 

All curves of different $f$ appear to cross each other in a Stokes number 
range of $7\lsim St_{h} \lsim 14$. In particular, at $St_{h} =12.4$, the variation of the normalized kernel is 
very small within $\lsim 10\%$ as $f$ changes from 0 to 1. 
We can thus assume that, for all values of $f$, the normalized kernels approximately 
cross at a common point around $St_{h} =12.4$ with $\Gamma^{\rm sph}/(\pi d^2) \simeq u' =6.8 u_\eta$.  
This means that, if the larger particle has a friction time $\tau_{\rm p,h} \simeq T_{\rm L} =14.4\tau_\eta$, 
the kernel per unit cross section may be estimated 
by the 1D rms flow velocity, $u'$, independent of the size of the smaller particle\footnote{
In the commonly-used prescription for particle 
collisions based on the model by V\"olk et al.\ (1980), the monodisperse kernel 
for $\tau_{\rm p} \simeq T_{\rm L}$ particles is about equal to the 3D rms 
flow velocity, $\sqrt{3} u'$, and thus overestimates  the collision rate by a factor 
of $\simeq \sqrt{3}$.}. 
The $f-$independence at $\tau_{\rm p,h} \simeq T_{\rm L}$ can be viewed as due to the 
cancelation of the dependences of the RDF and $\langle|w_{\rm r}|\rangle$ on $f$. 
At $St_{h} =12.4$, $\langle|w_{\rm r}|\rangle$ increases by a factor of 1.7 as $f$ goes from 1 to 0, 
while $g$ decreases by the same amount (Figs.\ \ref{rdf} and \ref{absv}). 

Considering that the motions of particles with $\tau_{\rm p} \simeq T_{\rm L}$ 
are insensitive to turbulent eddies at small scales, the constancy of the kernel with 
$f$ at $\tau_{\rm p,h} \simeq T_{\rm L}$ is likely independent of the Reynolds number, $Re$, 
of the flow. However, as pointed out in \S 2,  these particles couple to the driving scales 
of the turbulent flow, and their dynamics may be affected by how the flow is driven. 
Therefore, it remains to be confirmed whether the $f-$independence 
of the kernel at $\tau_{\rm p,h} \simeq T_{\rm L}$ is universal with respect to 
the flow driving mechanism.  For such a confirmation, one needs to conduct 
a new set of simulations with different driving schemes, 
which are computationally expensive and beyond the scope of the current work.

For $\tau_{\rm p,h} \gsim T_{\rm L}$, the kernel is larger at smaller $f$, 
because of the increase of $\langle|w_{\rm r}|\rangle$ with decreasing 
$f$. Interestingly, the opposite is seen for $1 \lsim St_{h} \lsim 12.4$, 
where the kernel is the smallest for $f=0$. As $f$ increases toward 
1, the clustering effect is stronger, and the increase of the RDF more than compensates the 
decrease of $\langle |w_{\rm r}| \rangle$, lifting all the kernels for $f>0$ above the black lines. 
The particle-tracer kernel (black lines) simply follows the scaling of $\langle|w_{\rm f,r}|\rangle$ (since $g =1$), 
and goes like $St_{h}^{1/2}$ (dotted black line segment) in the inertial range. 
At $f >0$, the RDF increases as $St_{h}$ decrease toward 1, and the increase is more rapid 
at larger $f$ (see Fig.\ \ref{rdf}). Thus the slope of the normalized 
kernel with $St_{h}$ in the inertial range becomes continuously shallower as $f$ increases. 
For particles of equal sizes in the inertial range, 
the opposite trends of $g$ and $\langle|w_{\rm r}|\rangle$ with $St$ (red lines in Figs.\ \ref{rdf} and \ref{absv}) lead 
to a very flat slope: the normalized kernel scales with $St_{h}$
roughly as $St_{h}^{0.15}$ (the red dotted-line segment). 
For $0<f<1$, the slope of the kernel is in between 0.5 and 0.15. For example, the measured 
slope is approximately 0.22,  0.3, 0.35, 0.4, and 0.43 for $f=\frac{1}{2}$, 
$\frac{1}{4}$, $\frac{1}{8}$, $\frac{1}{16}$, and $\frac{1}{32}$, respectively. 
These features are not captured by the kernel formulation commonly used in 
dust coagulation models, as clustering effect is neglected.  

In the right panel, we see that, for small particles with $St\lsim1$, 
the kernels at different $f$ almost cross at another common 
point with $St_{h} =0.6$ and $\Gamma_{\rm sph}/(\pi d^2) \simeq u_{\eta}$. 
Below $St_{h} =0.6$, the normalized kernel decreases as 
$f$ increases from 0 to 1, suggesting that, for small particles 
in the dissipation range, collisions of different particles are more 
frequent than between similar ones. 

The collision kernel commonly used in dust coagulation models ignores 
the effect of clustering and adopts the model of V\"olk et al.\ (1980) 
and its successors for the particle relative velocity. At a given $St_{h}$, the V\"olk-type 
models predict larger relative velocity between particles of different sizes than between 
particles of equal size (see, e.g., Ormel \& Cuzzi  2007). Therefore, the commonly-used 
kernel is smallest at $f=1$ and the largest at $f=0$ for any $St_{h}$. This trend is 
consistent with our results for $St_{\rm h}\lsim 1$ and $\tau_{\rm p, h} \gsim T_{\rm L}$, 
but is incorrect for $\tau_{\rm p, h}$ in the inertial range, where the clustering 
effect makes the kernel largest at $f=1$. 
Equation (28) of Ormel \& Cuzzi  (2007) for the rms relative velocity of inertial-range 
particles suggests that, in the commonly-used prescription,  the normalized collision 
kernel  for $f=0$ is larger by a factor of 1.5 than for $f=1$. 
On the other hand, our simulation data indicates that, in the inertial 
range, the kernel for $f=1$ is larger than the $f=0$ case by a factor of a few. 
As discussed in \S 3.5, in a realistic disk turbulence with a much larger Reynolds number than 
our simulated flow, the collision kernel for inertial-range particles of equal size ($f=1$) 
may exceed the kernel at $f\to 0$ by a significantly larger amount, if the RDF, $g$, has a 
significant Reynolds number dependence.

We also computed the cylindrical kernel $\Gamma^{\rm cyl}$ and found it almost coincides with 
$\Gamma^{\rm sph}$ except for small particles of similar sizes with $St_h \ll 1 $ 
(see \S 3.4). As mentioned earlier, the two formulations give equal estimate for the kernel if 
the direction of ${\bs w}$ is isotropic. In the cylindrical formulation, we can also split 
particle pairs of equal size into two types using a critical value, $|{\bs w}|_{\rm c}$, 
for the 3D relative velocity amplitude, $|{\bs w}|$. Particle pairs 
with $|{\bs w}| \ge |{\bs w}|_{\rm c}$ and $|{\bs w}| <|{\bs w}|_{\rm c}$ 
are counted as continuous and caustic pairs, respectively. With the 
choice of $|{\bs w}|_{\rm c} = 0.27 r/\tau_{\rm p}$, the caustic contribution 
to the cylindrical kernel per unit cross section, $g \int_{|{\bs w}|_{\rm c}}^{\infty} |{\bs w}| 
P( |{\bs w}|) d |{\bs w}|$, is $r-$independent and is almost equal to the caustic contribution 
from the spherical formulation.  Note that the critical value $|{\bs w}|_{\rm c} = 0.27 r/\tau_{\rm p}$ 
is different from the one used in the spherical formulation, and it is also different from 
the choice of $|{\bs w}|_{\rm c} = r/\tau_{\rm p}$ by Hubbard (2013).

We point out that there are uncertainties in the way to split the continuous 
and caustic pairs for small particles of equal size ($St\lsim1$, $f=1$). 
Theoretical models suggest that the critical relative velocity $w_{\rm c}$ (or $|{\bs w}|_{\rm c}$) 
scales as $\propto r/\tau_{\rm p}$, but do not predict its 
exact value. We selected a value $w_{\rm c}$ to obtain an $r-$independent 
contribution to the kernel, but there is no theoretical 
guarantee of the accuracy of our  choice for $w_{\rm c}$.  
It is possible that the $r-$independent caustic contribution we 
obtained at $r\gsim \frac{1}{4} \eta$ may not precisely 
represent the realistic kernel as $r\to 0$. In other words, as $r$ decreases 
below $\frac{1}{4} \eta$, the critical relative velocity $w_{\rm c}$ that minimizes 
the $r-$dependence could be different from the value we selected. 
Another uncertainty is that it is not clear whether, 
at the distance range we measured, the two types of pairs can be precisely 
divided simply by a single critical value for the relative velocity. 
The transition from continuous to caustic types may not occur sharply at a single 
relative velocity. Instead, the transition could be gradual, occurring over a relative velocity 
range, where particle pairs may not be definitely counted as continuous or caustic.  
Due to these uncertainties, the caustic kernel we obtained 
should be tested by future simulations that can measure the kernel at 
smaller scales, where the caustic contribution dominates.  

Despite the uncertainties, we think that the method of splitting two types of pairs is 
useful to provide physical insights to understanding the $r-$dependence of 
the collision kernel at $St\lsim 1$ and $f=1$ in the $r-$range explored.  
We also found that it gives a reasonable estimate for small 
equal-size particles at $r\to0$. Wilkinson et al.\ (2006) predict  that, for these 
particles, the normalized kernel at $r\to 0$ is given by $\propto \exp(-A/St)$, 
similar to an activation process. Our measured caustic kernel is in 
good agreement with this prediction using an activation threshold of $A \simeq 1.0$.

Hubbard (2012) argued that, in the monodisperse case, inertial 
particles that show significant clustering do not contribute to the collision rate, i.e.,
the particles that cluster are non-collisional.   
In our terminology, this argument essentially assumes that clustering is 
mainly caused by the continuous particle pairs, which do not contribute to 
the collision rate at $r\to0$, while the caustic pairs that contribute to the 
collision rate do not significantly contribute to clustering. 
This is likely true for particles of equal size with $St \lsim 1$. For these 
small particles, caustic pairs are very rare and make only a tiny fraction of 
the total number of particle pairs at a given small distance. 
Thus, the main contribution to clustering must be due to the 
rest of the particles, i.e., particle pairs of the continuous type. 
As discussed earlier, the continuous pairs do not 
contribute to the collision kernel, as the particle distance (or size), $r$, approaches zero 
(Wilkinson et al.\ 2006, Hubbard 2012). 

However, our data indicates that the argument of Hubbard (2012) is invalid 
for inertial-range particles with $ \tau_\eta \lsim \tau_{\rm p} \lsim T_{\rm L}$. 
In Paper I, we showed that caustic pairs start to dominate at $St \gsim 1$, 
and, at $St \gsim 3.11$, most particle pairs belong to the caustic type. Therefore, if 
significant clustering occurs for inertial-range particles, it must be from caustic 
pairs. Fig.\ \ref{rdf} shows that the RDF, $g$, is significantly above unity 
for inertial-range particles of equal size, suggesting that caustic 
particles do contribute to clustering. In fact, it is the clustering of the caustic pairs 
that makes the normalized kernel for particles of equal size in the inertial range 
larger than that between particles of different sizes. Thus, for inertial-range 
particle of equal size, it is not true that the clustering particles do not collide.   

For particle of different sizes, the RDF, the relative velocity and hence 
the kernel all converge at $r\gsim \frac{1}{4} \eta$, and it is thus 
not necessary to split the pairs into the two types. In Fig.\ \ref{rdf}, we see that, for a Stokes ratio 
of $f \gsim \frac{1}{4}$, there is a moderate degree of clustering at $St_{h} \simeq 1$. 
Since the kernel for $f \gsim \frac{1}{4}$ and $St_{h} \simeq 1$ already converges 
at $r = \frac{1}{4} \eta$, all particle pairs at that distance contribute to the collision rate. 
Therefore, clustering at $f \gsim \frac{1}{4}$ and $St_{h} \simeq 1$ would 
contribute to increase the collision rate, again in contrast to the argument of Hubbard (2012). 
In summary, the claim of Hubbard (2012, 2013) that the clustering 
particles do not contribute to the collision rate is not valid in general and applies only for the special case of 
small equal-size particles with $St \lsim1$. 

\subsection{Comparing with Commonly-used Kernel Formula}   

The kernel formula commonly used in dust coagulation models 
is $\Gamma^{\rm com} = \pi d^2 \langle w^2\rangle^{1/2}$, where 
the rms relative velocity, $\langle w^2\rangle^{1/2}$, is usually taken from 
the model of V\"olk et al.\ (1980) or its successors. 
$\Gamma^{\rm com}$ is in a simpler form than $\Gamma^{\rm sph, cyl}$, 
and in this section we use our simulation data 
to assess the accuracy of the simplified formula for the collision kernel. We 
defer a systematic test of  V\"olk-type models for $\langle w^2\rangle^{1/2}$ to a later work.  	

\begin{figure*}[t]
\includegraphics[height=3.1in]{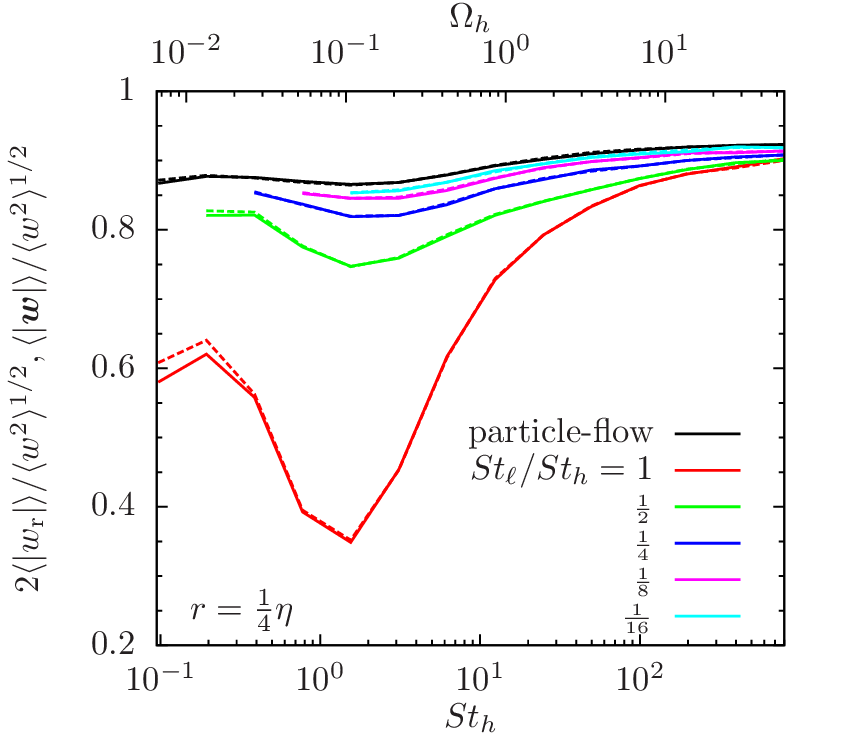}
\includegraphics[height=3.1in]{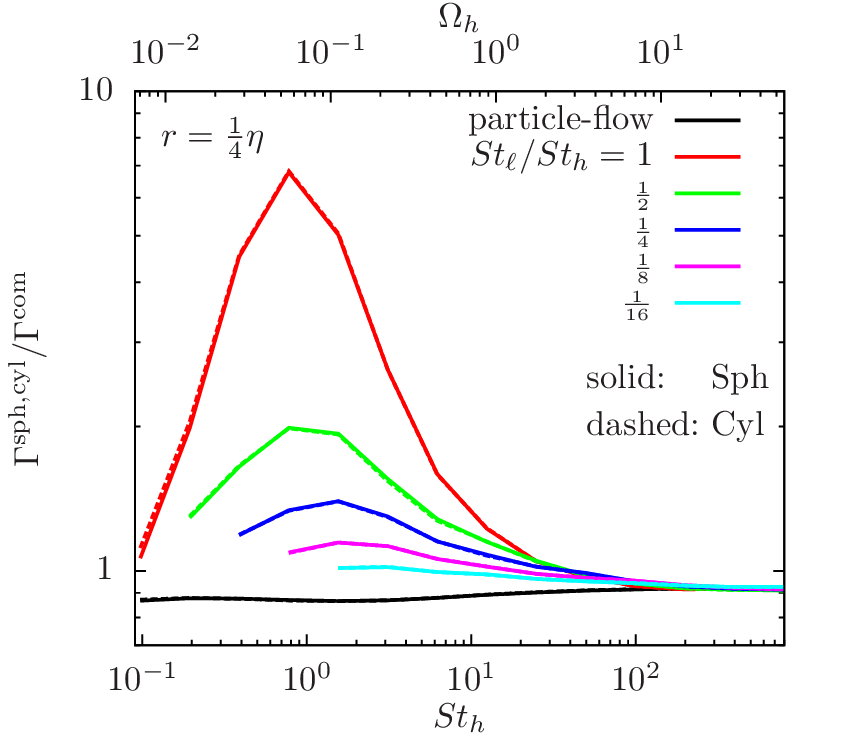}
\caption{Left panel: the mean to rms ratios, $C_{12}^{\rm sph} \equiv 2\langle |w_{\rm r}|\rangle/ \langle w^2\rangle^{1/2} $ 
(solid) and $C_{12}^{\rm cyl}  \equiv \langle |{\bs w}| \rangle/\langle w^2\rangle^{1/2}$ (dashed), at $r=\frac{1}{4}\eta$.
Right panel: The ratios of  spherical (solid)  and cylindrical (dashed) kernels to the commonly-used formulation, 
$\Gamma^{\rm com} = \pi d^2 \langle w^2\rangle^{1/2}$.  The effect of turbulent clustering is not accounted for in 
the commonly-used kernel, $\Gamma^{\rm com}$.
The kernels are measured at $r=\frac{1}{4} \eta$.
}
\label{testkernel} 
\end{figure*}

In comparison to the spherical and cylindrical kernels $\Gamma^{\rm sph, cyl}$, $\Gamma^{\rm com}$ 
neglects the clustering effect, 
and uses the 3D rms of the relative velocity rather than 
the mean relative velocity, $\langle |w_{\rm r}| \rangle$ or $\langle |{\bs w}| \rangle$.  
The ratios of $\Gamma^{\rm sph, cyl}$ to $\Gamma^{\rm com}$ 
can be written as $g C_{12}^{\rm sph, cyl}$, 
where $C_{12}^{\rm sph} \equiv 2\langle |w_{\rm r}|\rangle/ \langle w^2\rangle^{1/2}$ 
and $C_{12}^{\rm cyl} \equiv \langle |{\bs w}| \rangle/\langle w^2\rangle^{1/2}$.  
The ratios $C_{12}^{\rm sph, cyl}$ compare the 
mean and rms relative speeds, corresponding to the first- and second- order 
moments of ${\bs w}$, respectively. In the left panel Fig.\ \ref{testkernel}, we show 
$C_{12}^{\rm sph}$ (solid) and $C_{12}^{\rm cyl}$ (dashed) computed from our simulation 
data at $r= \eta/4$. For all Stokes pairs, the ratios are smaller than unity, 
meaning that using the rms relative velocity tends to overestimate the kernel. 
Except for a small difference of $5\%$ at $St_h \ll 1$ and $f=1$, the solid lines 
almost coincide with the dashed ones, suggesting that $2 \langle| w_{\rm r}\rangle = \langle |{\bs w}| \rangle$, 
and hence $\Gamma^{\rm sph} = \Gamma^{\rm cyl}$, within an uncertainty 
of only 5\%.  The near equality of  $\Gamma^{\rm sph} \simeq 
\Gamma^{\rm cyl}$  was expected under the assumption of isotropy for the 
direction of ${\bs w}$.

The ratios, $C_{12}^{\rm sph, cyl}$, depend on the shape of the probability 
distribution function of ${\bs w}$, particularly on the central part of the PDF, 
as both the mean and the rms are low-order moments.  If ${\bs w}$ is Gaussian, 
the PDF of the 3D amplitude $|{\bs w}|$ is Maxwellian, and it is easy to show 
$C_{12}^{\rm cyl} = ({8/3\pi})^{1/2} \simeq 0.92$. The same value is expected 
for $C_{12}^{\rm sph}$. As the radial rms relative speed, $\langle w_{\rm r}^2\rangle^{1/2}$, 
is typically smaller than the 3D rms, $\langle w^2\rangle^{1/2}$, by a 
factor of $\sqrt{3}$ (see Paper II), we have $C_{12}^{\rm sph} \simeq (4/3)^{1/2} \langle |w_{\rm r}|\rangle/\langle w_{\rm r}^2\rangle^{1/2}$. 
With a Gaussian distribution for $w_{\rm r}$, $\langle |w_{\rm r}|\rangle/\langle w_{\rm r}^2\rangle^{1/2} = ({2/\pi})^{1/2}$, 
and thus $C_{12}^{\rm sph}$ is also equal to $(8/3\pi)^{1/2}$. A value of $\simeq 0.92$ 
for $C_{12}^{\rm sph, cyl}$ is observed in the left panel of Fig.\ \ref{testkernel} at $\tau_{\rm p,h} \gg T_{\rm L}$, 
where the PDF of ${\bs w}$ is approximately Gaussian (see Paper III). 
 
As discussed in Paper I, the mean to rms ratios are smaller if the PDF of ${\bs w}$ 
is non-Gaussian with a sharper inner part and fatter tails (see Fig.\  16 of Paper I). 
For particles of equal size with $\tau_{\rm p} \lsim T_{\rm L}$, the PDF is highly 
non-Gaussian with very fat tails, and the ratios $C_{12}^{\rm sph, cyl}$ are significantly 
smaller than 0.92. In particular, the degree of non-Gaussianity peaks at $St \simeq 1$, 
leading to a dip in the ratio at $St \simeq 1$, where $C_{12}^{\rm sph, cyl} \simeq 0.3-0.4$. 
In Paper III, we showed that the non-Gaussianity of 
the relative velocity PDF decreases with decreasing Stokes ratio $f$, 
and this is responsible for the increase of  $C_{12}^{\rm sph, cyl}$ as $f$ decreases toward 0. 
In the case of $f=0$, the inner part of the PDF for the particle-flow relative velocity 
is nearly Gaussian for all particles (see Fig.\ 1 of Paper III), 
which explains why the black lines for $f=0$ are close to 0.9 at all $St$. 
The left panel of Fig. \ref{testkernel} suggests  that using  the rms instead of the mean 
relative velocity in the collision kernel tends to overestimate the collision rate, 
especially for particles of similar sizes.     

We also computed $C_{12}^{\rm sph, cyl}$ at other distances, $r$.  
For $f\lsim \frac{1}{2}$, the ratio converges at $r=\frac{1}{4} \eta$, 
corresponding to the convergence of the PDF shape found 
in \S 4.2 of Paper III. But for particles of equal size with $St \lsim 6$, $C_{12}^{\rm sph, cyl}$ 
have not converged and decrease with decreasing $r$ due to the fatter 
PDF shape at smaller $r$ (Paper I). For these particles, the ratios 
at $r\lsim \eta/4$ could be significantly smaller than the values shown in the left panel of Fig.\ \ref{testkernel}.

In Fig.\ \ref{testkernel}, we plot the ratio, $\Gamma^{\rm sph, cyl}/\Gamma^{\rm com}$, 
in our simulation. In the calculation of $\Gamma^{\rm com}$, we 
used the 3D rms relative velocity measured from our data. Again, 
the  near coincidence of the solid and dashed lines for $\Gamma^{\rm sph}$ and $\Gamma^{\rm cyl}$
indicates that the two formulations give similar estimates 
for the collision rate. 
In the limit $f\to 0$ (black lines), the ratio of the commonly-used 
formula to $\Gamma^{\rm sph, cyl}$ is about 0.9 at all $St_{h}$, meaning that it 
overestimates the kernel slightly by $\lsim 10\%$. For this particle-tracer 
case, the RDF $g$ is unity, and the overestimate is 
because $\langle w^2\rangle^{1/2}$ is larger than $2 \langle| w_{\rm r}\rangle$ by 10\% (see the 
left panel). For $f\lsim \frac{1}{4}$, $\Gamma^{\rm com}$ is a good 
approximation to the collision kernel except for an overestimate of $\lsim 40\%$ at $St_h\simeq1$.

For particles of similar sizes ($f=1$ and $\frac{1}{2}$), the ratios of 
$\Gamma^{\rm sph, cyl}$ to $\Gamma^{\rm com}$ are close to unity for $\tau_{\rm p} \gsim T_{\rm L}$, 
but become  significantly larger as $St_{h}$ decreases in the 
inertial range toward $\simeq 1$. This means that $\Gamma^{\rm com}$ underestimates the 
kernel for particles of similar sizes in the inertial range and in particular at $St_h \simeq 1$.   
The ratios $\Gamma^{\rm sph, cyl}/\Gamma^{\rm com}$ converge at $r=\frac{1}{4} \eta$ 
except for $f=1$. For equal-size particles, $\Gamma^{\rm sph, cyl}/\Gamma^{\rm com}$ 
keeps increasing as $r$ decreases to $\frac{1}{4} \eta$, and the 
underestimate of $\Gamma^{\rm com} $ would be more severe at  $r<\frac{1}{4} \eta$.
In general, how $\Gamma^{\rm com}$ compares to $\Gamma^{\rm sph, cyl}$ depends on two opposite effects: 
using the rms relative velocity instead of the mean tends to overestimate the 
collision rate, while neglecting the clustering effect causes an underestimate. 
Both effects are strongest for particles of similar sizes at $St_{h} \simeq 1$ 
(see the left panel of Fig.\ \ref{testkernel} and Fig.\ \ref{rdf}), 
but it turns out that the second effect dominates for $St_h$ in 
the inertial range, where $g$ is significantly larger than $1/C_{12}^{\rm sph,cyl}$. 
Another interesting consequence of neglecting clustering is that, as $f$ decreases from 1 to 0, 
$\Gamma^{\rm com}$ increases because $\langle w^2\rangle$ is larger at smaller $f$ (see Fig.\ 7 of paper II). 
This is in contrast to the result shown in the right panel of Fig.\ \ref{kernel}, 
where the kernel decreases with decreasing $f$ 
for $St_h$ in the inertial range. 

In summary, the commonly-used kernel formula provide a good approximation for 
very different particles with the Stoke ratio $f\lsim \frac{1}{4}$ or if the larger particle 
has $\tau_{\rm p,h} \gsim T_{\rm L}$. 
However, it significantly underestimates the collision rate for particles of 
similar sizes (with $f\gsim \frac{1}{2}$) with $\tau_{\rm p,h} \lsim T_{\rm L}$, 
especially for $St_{\rm h} \simeq 1$.

\subsection{Extrapolation to Realistic Reynolds Numbers}

Due to the limited numerical resolution, the Reynolds number, $Re$, 
of our simulated flow is only  $\simeq 10^3$, much smaller than 
the typical value of $Re \simeq 10^6-10^8$ in protoplanetary disks. 
To study dust particles in protoplanetary turbulence, the results of our 
simulation should be appropriately extrapolated to a realistic $Re$. 
An accurate extrapolation requires the $Re-$dependence 
of the collision kernel, which is currently unknown. Here we 
offer some speculations on how to extrapolate the measured 
results to higher $Re$. Studies of the $Re-$dependence using 
higher-resolution simulations will be conducted in future work.

In our simulation, the inertial range is short, and the timescale separation 
between the driving  scale and the Kolmogorov scale 
is only one order of magnitude. More precisely, we have 
$T_{\rm L}/\tau_\eta \simeq 14.4$ in our flow. 
Although we count all particles with $\tau_\eta \lsim \tau_{\rm p} \lsim T_{\rm L}$ 
inertial-range particles, strictly speaking, there is only one species of 
particles in our simulation that are definitely not affected by 
either the dissipation or the driving scales of the flow (Paper III). 
This species is particles with $\tau_{\rm p} = 6.21 \tau_{\eta}$, 
or equivalently $\tau_{\rm p} = 0.43 T_{\rm L}$. 
Particles with $\tau_{\rm p} \gsim 0.43 T_{\rm L}$ are coupled to eddies 
significantly above the dissipation range, and their dynamics is insensitive 
to the smallest eddies in the flow. Thus, when normalized to the flow 
properties at driving scales, the collision statistics of these particles is expected 
to be independent of $Re$. 
Our measured kernel for these particles can be directly applied to corresponding 
particles with $\Omega  \equiv \tau_{\rm p}/T_{\rm L} \gsim 0.43$ in protoplanetary disks.  
Therefore, we only need extrapolations for $\Omega \lsim 0.43$ particles 
in the inertial and dissipation ranges of the real flow. 

As $Re$ increases, the separation between $\tau_\eta$ and $T_{\rm L}$ 
increases as $Re^{1/2}$. Considering that $Re\simeq10^3$ and 
$T_{\rm L}/\tau_\eta \simeq 14.4$ in our flow, we have $\Omega \simeq 2.2 Re^{-1/2}St$ 
for larger $Re$. This suggests that, if the driving of the flow is 
fixed, then, with increasing $Re$, $St=1$ particles correspond to 
smaller size with $\Omega \simeq 2.2 Re^{-1/2}$. 
Particles with $2.2 Re^{-1/2} \lsim \Omega \lsim 1$ lie in the inertial range, 
and we need to extrapolate the kernel down to $\Omega \simeq  2.2 Re^{-1/2}$. 

A simple extrapolation is to assume that the measured slope for the normalized 
kernel in the inertial range (see Fig.\ \ref{kernel}) is independent of the Reynolds 
number. With this assumption, the normalized kernel for equal-size particles ($f=1$) 
in the inertial range is approximately given by $\simeq u' \Omega^{0.15}$, while 
in the $f \to 0$ limit it scales as $\simeq u' \Omega^{1/2}$ (see Fig.\ \ref{kernel} and \S 3.3).
In this case, the effect of clustering significantly enhances the kernel for 
all inertial-range particles of similar sizes.  Interestingly,  with this extrapolation, we find that, 
 for a realistic value of $Re \gsim10^6-10^8$, the normalized kernel for equal-size particles at 
 $St_h =1$ (corresponding to $\Omega_{\rm h} = 2.2 \times 10^{-3} - 2.2 \times 10^{-4}$) 
 is larger by a factor of 10-20 than the particle-tracer limit with $f\to0$. This suggests 
significantly higher collision frequency between similar particles than 
between particles of different sizes, as the particles just grow into the inertial range. 
Note that, in our flow with low $Re$, the variation of the kernel 
at $St_{\rm h} =1$ is quite small, increasing only by a factor 
of $\le 2$ as $f$ increases from $0$ to $1$ (see Fig.\ \ref{kernel}).

In the extrapolation above for equal-size particles, an 
implicit assumption was made for the scaling of the 
RDF with $\Omega$ in the inertial range. Since the relative 
velocity is expected to scale as $\propto \Omega^{1/2}$ (or $St^{1/2}$) 
in the inertial range\footnote{A variety of models predict an $\Omega^{1/2}$ 
scaling for the rms relative velocity in the inertial range. 
A similar scaling may be expected for $\langle |w_{\rm r}| \rangle$. However, the scaling of $\langle |w_{\rm r}| \rangle$ 
could be steeper than $\Omega^{1/2}$, as the ratio $\langle |w_{\rm r}| \rangle/\langle w^2 \rangle^{1/2}$ 
decreases with decreasing $\Omega$ or $St$ in the inertial range (see the left panel of Fig.\ \ref{testkernel}).}, 
the assumed $\Omega^{0.15}$ scaling for the normalized kernel, $\Gamma^{\rm sph}/\pi d^2$ 
($\equiv 2 g \langle |w_{\rm r}|\rangle$), holds only if the RDF increases with decreasing $\Omega$ as $\propto\Omega^{-0.35}$ in the inertial range. 
The RDF for $f=1$  in our flow does not have a clear 
power-law range with $\Omega$ (see Fig.\ \ref{rdf}). As mentioned 
earlier, strictly only one species of particles ($St=6.21$) lie 
in the inertial range in our flow, and it is possible that 
$g$ may show an extended power-law scaling with $\Omega$ 
in a high$-Re$ flow with a wider range of scales unaffected 
by the dissipation or the driving scales. The possibility, 
however, needs to be verified by simulations at higher resolutions.

If the scaling $g \propto \Omega^{-0.35} $ is valid for particles in the 
inertial range, it would imply a $Re-$dependence of the RDF for particles 
at fixed Stokes numbers, $St$. Inserting $\Omega \propto Re^{-1/2}St$ to 
this scaling gives $g (St \simeq 1) \propto Re^{0.175}$ for $St \simeq 1$ 
particles in turbulent flows of different $Re$. The $Re-$dependence 
of the RDF at fixed Stokes numbers around or below unity have 
been studied in the literature, but, to our knowledge, a conclusive answer 
is still lacking. For example, the simulation of  Collins \& Keswani (2004) 
indicated that the RDF of $St\simeq 1$ particles converges already at $Re\simeq 600$ 
(see also Hogan \& Cuzzi 2001), while Falkovich \& Pumir (2004) found 
that the clustering intensity shows a significant increase with increasing $Re$. 

Clearly, if the $Re$-dependence of the RDF at $St \simeq 1$ is 
weaker than $Re^{0.175}$,  the increase of $g$ with decreasing $\Omega$ 
would be slower than $\Omega^{-0.35}$, and thus the scaling of the 
normalized kernel for $f=1$ would be steeper than $\Omega^{0.15}$.  
An extreme case is that the RDF for $St $ fixed around $\simeq 1$  
is independent of $Re$. In that case, the RDF for inertial-range particles
with $\tau_{\rm p}$ significantly away from the dissipation scales can only vary 
slowly with $\Omega$, and, for these particles, the normalized kernel would  
scale as $\Omega^{1/2}$, as it just follows scaling of the relative 
velocity. This scaling is similar to the $f\to 0$ limit.   
Therefore, we would expect the kernels for $f=1$ and $f\to 0$ to 
remain close to each other as $\Omega_{h}$ decreases in the inertial range. 
In the inertial range, clustering does not considerably increase the collision rate of similar-size particles.  
Only as $St_h$ approaches 1, would the monodisperse RDF start to increase significantly, 
which would tend to increase the kernel for $f=1$ above the $f\to0$ case by a factor of 2 or so. 
This behavior is in contrast to the extrapolation based on the slope of 
the normalized kernel measured in our simulation.

In summary, the extrapolation of the measured kernel from our simulation 
to a realistic flow with much larger $Re$ is uncertain. 
If we assume that the measured slope of the normalized 
kernel  at fixed $f$ can be applied to larger $Re$, then the decrease of the 
kernel for $f=1$ with decreasing $\Omega$ in the inertial range is slower than the limit $f \to 0$, and, 
at $\tau_{\rm p,h} \simeq \tau_\eta$, the collision rate between similar particles is much larger than between 
different particles.  On the other hand, if the RDF at a fixed Stokes number $St$ 
around unity is $Re-$independent, then the normalized kernels for $f =1$ and $f=0$ 
have a similar scaling with $\Omega_{h}$ and likely stay close to each other for 
any $\Omega_{h}$ in the inertial range.  
This question will be settled with future higher-resolution ($\ge 1024^3$) simulations, 
where the scalings of $g$ and of the normalized kernel with $\Omega$ (in the inertial range), and   
the $Re-$dependence of $g$ at fixed Stokes numbers can be checked.
     
\section{The Collision-Rate Weighted Statistics}

Coagulation models for dust particle growth need not only the collision kernel to 
calculate the collision rate, but also the collision velocity or collision energy 
to determine the collision outcome. While a small collision velocity favors particle 
sticking and growth, a large impact velocity leads to bouncing or destruction 
of the colliding particles. Until recently, dust coagulation models usually adopted a single collision velocity, 
typically the rms, to judge the collision outcome of two particles of given sizes. 
However, due to the stochastic nature of turbulence, the collision 
velocity for each size pair is not single-valued, but has a 
probability distribution. Thus, even for particles of exactly the 
same properties, the collision outcome can be different, and the 
probability distribution function (PDF) of the collision 
velocity is needed to calculate the fractions of collisions resulting in sticking, bouncing 
and fragmentation. Windmark et al.\ (2012) and Garaud et al.\ (2013) 
have shown that accounting for the collision velocity PDF makes a significant  
difference in the predicted particle size evolution in protoplanetary disks. 
 
Garaud et al.\ (2013) emphasized that the collision velocity PDF 
to be used in a coagulation model should include a collision-rate weighting 
factor to account for the higher collision frequency of particle 
pairs with larger relative velocity. The importance of collision-rate weighting was also pointed 
out by Hubbard (2012), who argued that the commonly-used rms relative 
velocity with equal weight for each nearby particle pair is not appropriate for 
applications to dust particle collisions. A collision-rate weighted rms was proposed by 
Hubbard (2012). In Papers II and III, we have systematically studied 
the unweighted rms  and PDF of turbulence-induced relative velocity. 
In this section, we compute the collision-rate weighted statistics from 
our simulation data. 
   
In the cylindrical kernel formulation, the collision rate is $\propto|{\bs w}|$, 
and the collision-rate weighted distribution of the 3D amplitude can be obtained from the 
unweighted PDF, $P(|\bs {w}|)$, simply by $P_{\rm cyl} (|\bs {w}|) =  |\bs {w}| P(|\bs {w}|)/\langle |\bs {w}|\rangle$ 
(see eq.\ (36) in Garaud et al.\ 2013).
The weighted PDF of $|\bs{w}|$ in the spherical formulation is more 
complicated, and in terms of the PDF, $P_{\rm v}(\bs {w})$ of  the vector ${\bs w}$, it can be defined  as, 
\begin{gather}
P_{\rm sph} (|\bs {w}|) = -\frac{1}{N}\int\limits_{\pi/2}^{\pi} d\theta
\int\limits_0^{2\pi} d\phi  \sin(\theta)  \cos(\theta) |{\bs w}|^3P_{\rm v}(|\bs {w}|, \theta, \phi),  
\label{sphpdf}
\end{gather}
where a weighting factor  $-w_{\rm r}$($= |{\bs w}| \cos(\theta)$) is applied and  the integration limits for $\theta$ 
count only approaching pairs with negative $w_{\rm r}$. The normalization 
factor is $N = -\int_{-\infty}^{0} w_{\rm r} P(w_{\rm r}) dw_{\rm r}$, with $P(w_{\rm r})$ the PDF of the radial component.  

If the direction of ${\bs w}$ is isotropic relative to ${\bs r}$, 
we have $P_{\rm v}(|{\bs w}|, \theta, \phi) = P_{\rm v}(|{\bs w}|)$, 
and a calculation of  the double integral in 
eq.\ (\ref{sphpdf}) gives $-\pi |{\bs w}|^3P_{\rm v}(|\bs {w}|)$. 
Using eq.\ (\ref{wrpdf}) for $P(w_{\rm r})$, it is easy to show that $N = \pi \int_0^{\infty} |{\bs w}|^3 P_{\rm v} (|{\bs w}|) d  |{\bs w}|$.  
We thus have $P_{\rm sph} (|\bs {w}|) = |{\bs w}|^3P_{\rm v}(|\bs {w}|)/\int_0^{\infty} |{\bs w}|^3 P_{\rm v} (|{\bs w}|) d  |{\bs w}| $, 
which is identical to $P_{\rm cyl} (|\bs {w}|)$. Therefore, the weighted PDFs of the 3D amplitude, 
$|{\bs w}|$, with the spherical and cylindrical formulations are equal under the isotropy assumption for 
${\bs w}$.

We measure $P_{\rm sph} (|\bs {w}|)$ from our simulation data as 
$P_{\rm sph} (|\bs {w}|) =  \sum_{i} H (-w_{\rm r}^{(i)}) w_{\rm r}^{(i)} \delta (|\bs {w}| -|\bs {w}|^{(i)}) 
/ \sum_i H (-w_{\rm r}^{(i)}) w_{\rm r}^{(i)} $, where the sum is over all 
particle pairs available at a given small distance and $H$ is the Heaviside 
step function to exclude the separating particle pairs moving away 
from each other. The PDF measured this way is consistent with $P_{\rm sph} (|\bs {w}|)$ defined by eq.\ (\ref{sphpdf}).

\subsection{The Collision-Rate Weighted RMS}

We start by considering the collision-rate weighted rms relative velocity, $w'_{\rm sph}$ 
and $w'_{\rm cyl}$, in the two formulations. In the cylindrical formulation, $w'_{\rm cyl} \equiv (\int_0^{\infty} |{\bs w}|^2 P_{\rm cyl} (|{\bs w}|) d |{\bs w}|)^{1/2} 
= (\langle |{\bs w}|^3 \rangle/ \langle |{\bs w}| \rangle)^{1/2}$ (see Hubbard 2012). 
Similarly, we have $w'_{\rm sph} \equiv (\int_0^{\infty} \bs {w}^2P_{\rm sph} (|\bs {w}|) d|\bs {w}|)^{1/2}  
 =   (\langle w_{\rm r}{\bs w}^2 \rangle_-/ \langle  w_{\rm r}\rangle_-)^{1/2}$ for the spherical formulation. 
The subscript ``-" in the ensemble averages indicates that only approaching particle pairs with $w_{\rm r} <0$ 
are counted, e.g., $\langle  w_{\rm r}\rangle_- \equiv \int_{-\infty}^{0} w_{\rm r}   P(w_{\rm r}) dw_{\rm r}/\int_{-\infty}^{0} 
P(w_{\rm r}) dw_{\rm r}$. With collision-rate weighting, the  rms relative velocity, $w'_{\rm sph}$ 
and $w'_{\rm cyl}$, provides a measure for the average collisional energy per collision. 
Although the rms itself is not sufficient for modeling  collisional  growth of dust particles,
a calculation of $w'_{\rm sph}$ and $w'_{\rm cyl}$ helps 
illustrate the effect of the collision-rate weighting. Also,
 the collision-rate weighted rms is likely more appropriate 
than the unweighted one to use in coagulation models that 
ignore the distribution of the collision velocity.  

\begin{figure*}[t]
\includegraphics[height=2.9in]{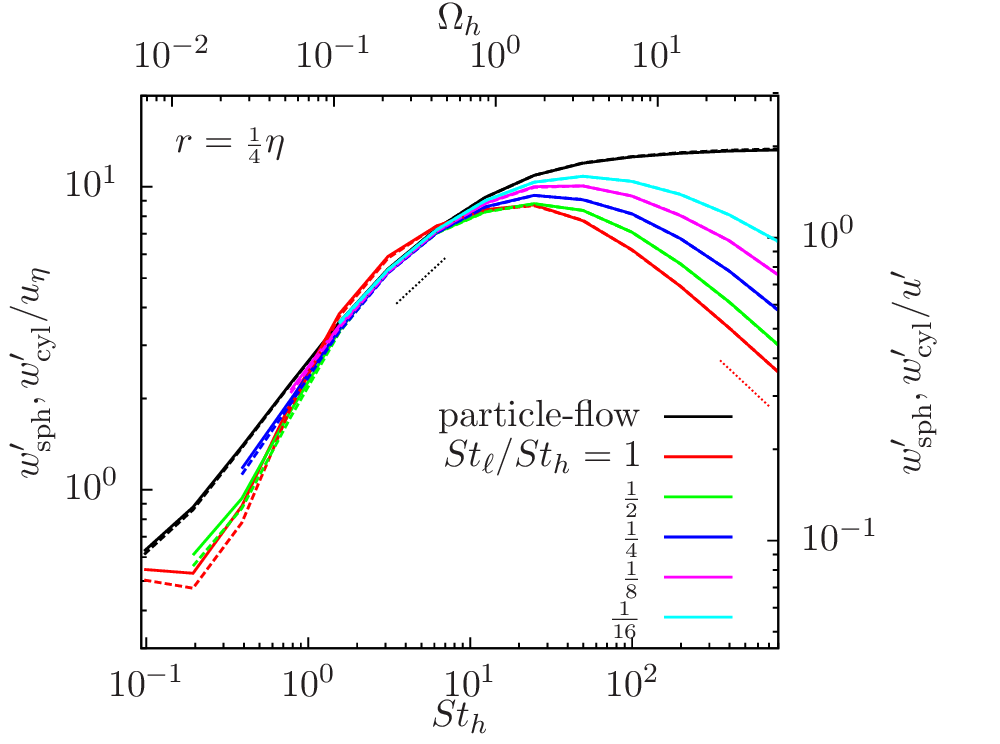}
\includegraphics[height=2.9in]{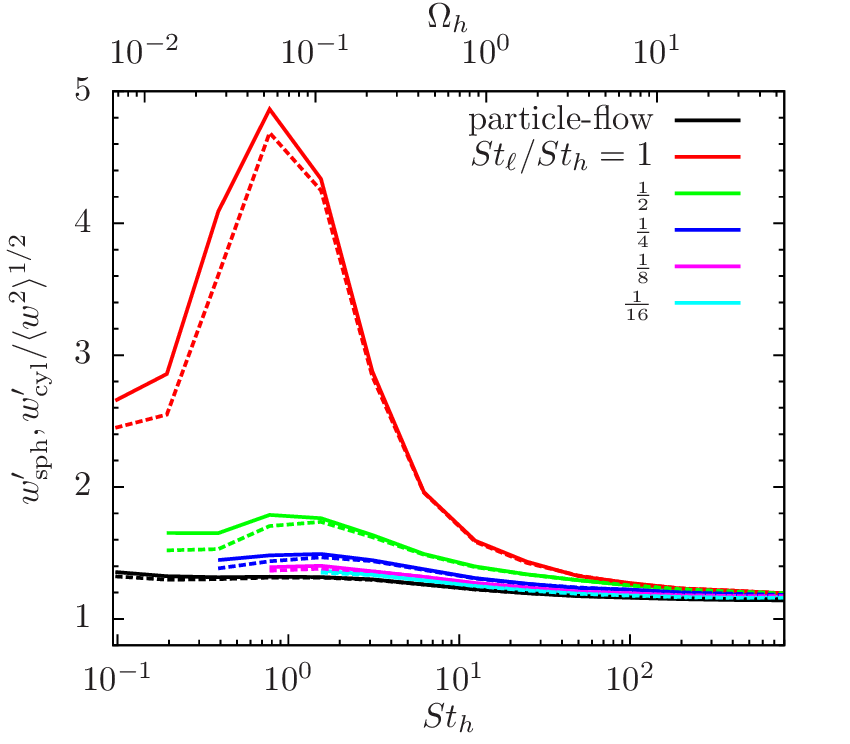}
\caption{Left panel: collision-rate weighted rms relative velocity in the spherical (solid, $w_{\rm sph}^{\prime}$) 
and cylindrical (dashed, $w_{\rm cyl}^{\prime}$) formulations at $r=\frac{1}{4} \eta$. 
Dotted line segments denote $St_{h}^{1/2}$ and   $St_{h}^{-1/2}$ scalings. Right panel: ratios 
of $w_{\rm sph}^{\prime}$ and $w_{\rm cyl}^{\prime}$ to the unweighted rms.
}
\label{collisionv} 
\end{figure*}

Fig.\ \ref{collisionv} shows results for $w'_{\rm sph}$ (solid) and $w'_{\rm cyl}$ (dashed) 
measured at $r=\frac{1}{4} \eta$. The solid and dashed lines in Fig.\ \ref{collisionv} almost 
coincide. The equality of the weighted rms relative velocity in the two formulations was expected from the 
isotropy of ${\bs w}$. Slight differences ($\lsim 10\%$) are found at small $St_{h}$ and $f \gsim 1/2$, 
because, for these Stokes pairs, the assumption of perfect isotropy 
of ${\bs w}$ is not satisfied.   

Interestingly, if the larger particle is in the inertial range, i.e., 
$\tau_{\eta} \lsim \tau_{\rm p,h} \lsim T_{\rm L}$, $w'_{\rm cyl}$ 
and $w'_{\rm sph}$ are almost independent of $f$, or the friction time, $\tau_{\rm p,l}$, 
of the smaller particle. In Paper II, we showed that the unweighted rms relative velocity, 
$\langle {\bs  w}^2 \rangle^{1/2}$, increases with decreasing $f$ because 
the generalized acceleration contribution to the relative velocity, corresponding 
to different responses of different particles to the flow velocity,  increases 
with increasing fiction time difference (see Fig.\ 7 of Paper II). 
To understand the invariance of $w'_{\rm sph, cyl}$ with $f$ for 
$St_h$ in the inertial range, it is helpful to compute the ratio 
of $w'_{\rm sph, cyl}$ to the unweighted rms, $\langle {\bs  w}^2 \rangle^{1/2}$, which is shown 
in the right panel of Fig.\ \ref{collisionv}.  The ratio depends on the PDF shape 
of ${\bs w}$. Since $w'_{\rm sph}$ and $w'_{\rm cyl}$ correspond to higher (3rd) 
order statistics than the unweighted rms, the ratio $C_{32} \equiv w'_{\rm sph, cyl}/\langle {\bs  w}^2 \rangle^{1/2}$ 
is larger if the PDF of ${\bs w}$ has fatter tails. 
In Paper III we found that, at a given $St_{h}$, the PDF shape of the relative velocity 
is the fattest for  particles of equal size ($f=1$), and becomes continuously
thinner as $f$ decreases. This explains why the ratios shown in the right 
panel increase as $f$ increases from 0 to 1.  
For $St_{h}$ in the inertial range, the larger ratio $C_{32}$ at larger $f$ 
happen to approximately cancel out the decrease of the unweighted rms, 
$\langle {\bs  w}^2 \rangle^{1/2}$, with increasing $f$ (see Fig.\ 7 of Paper II), 
leading to almost invariance of $w'_{\rm sph, cyl}$ with $f$. This $f-$independence simplifies 
the estimate of the average collision energy of inertial-range particles. 
For $\tau_{\rm p,h} \simeq T_{\rm L}$, the average collision velocity for 
each collision is found to be $\simeq1.3 u'$, 
independent of $f$. The collision velocity roughly follows a $St_{h}^{1/2}$ 
scaling for $St_{h}$ in the inertial range, which, however, 
needs to be verified by larger simulations with a broader inertial range.

For $St_h \lsim 1$ and $\tau_{\rm p, h} \gsim T_{\rm L}$, the increase of $C_{32}$ with 
increasing $f$ is not sufficient to compensate the decrease of $\langle {\bs  w}^2 \rangle^{1/2}$, 
and the weighted rms is smaller at larger $f$.  At $f =1$, the ratio $w'_{\rm sph, cyl}/\langle {\bs  w}^2 \rangle^{1/2}$ 
strongly peaks at $St \simeq 1$, corresponding to the result of Paper III that the PDF is the 
fattest at $St \simeq 1$.  
As shown in Paper III, the PDF of  ${\bs w}$ becomes nearly Gaussian for sufficiently 
large $\tau_{\rm p,h}$ ($\gg T_{\rm L}$), and, for a Gaussian distribution, it is easy to see 
that $w'_{\rm cyl}/\langle {\bs  w}^2 \rangle^{1/2} = \sqrt{4/3} \simeq 1.15$. A ratio of 
1.15 is actually observed at large $St_{h}$ in the right panel of Fig.\ \ref{collisionv}. 
The right panel of Fig.\ \ref{collisionv} also suggests that using the unweighted rms 
$\langle {\bs  w}^2 \rangle^{1/2}$ significantly underestimates the average collision 
velocity per collision, especially for the collisions between 
particles of similar size with $St \simeq 1$.

For $f\le{1/2}$, the measured rms velocity with collision-rate weighting 
converges at $r=\frac{1}{4} \eta$. The same is found for equal-size 
particles with $St \gsim 0.8$. However, complexities arise for 
smaller particles of equal size with $St \lsim 0.4$. For these 
particles, the weighted rms collision velocity has not 
reached convergence at $r = \frac{1}{4} \eta$, and a refinement 
is needed before it can be used in practical applications.   
In \S 3.3, in order to estimate the kernel in the $r\to 0$  limit, we attempted 
to isolate an $r-$independent kernel contribution by separating out the caustic pairs. 
However, we find that the method does not apply here to estimate 
the weighted rms collision velocity at $r\to0$ or to obtain an $r-$independent 
contribution to the rms. 

The idea of selecting the caustic pairs for small particles of 
equal size with $St \ll 1$ is to approximate the collision statistics 
at $r\to 0$ by the measured statistics of the relative velocity of particle pairs at a finite 
$r$ above a threshold. Although it was useful to obtain an $r-$independent 
kernel contribution, this approximation is not justified in general. 
When a threshold relative velocity is applied to select caustic pairs, 
the continuous pairs with low relative velocity are excluded, 
which pushes the rms relative velocity to higher values than the 
overall rms including all pairs at a given $r$. 
However, the rms collision velocity at $r\to 0$ is expected 
to be smaller than the overall rms at finite $r$, because 
the overall rms decreases with decreasing $r$. 
This suggests that the rms collision velocity of the caustic 
pairs does not correctly represent the rms in the $r\to 0$ limit. Thus, 
the relative velocity PDF excluding the continuous pairs 
at a finite $r$ is not equivalent to the PDF in the $r\to 0$ 
limit; instead it overestimates the collision velocity at 
$r\to0$. We note that, when computing  the rms collision velocity in 
simulations, Hubbard et al.\ (2012) proposed to count 
particle pairs with relative velocity above a threshold only. Based on the 
above discussion, there is uncertainty in the rms collision velocity 
measured this way as an estimate for $r\to0$, and some care 
is needed when using such a method.
   
It turns out that isolating the caustic pairs 
from the continuous pairs does not help the convergence 
of the rms collision velocity.  
The threshold relative velocity is larger at larger $r$, 
as it is chosen to be linear with $r$ (see \S 3.3).  
It turns out that, with a stronger threshold, 
the weighted rms relative velocity of caustic pairs at 
a larger $r$ exceeds the overall rms by a larger amount than at smaller $r$.  
This means that the convergence of the rms relative velocity of caustic pairs is 
actually slower than the overall rms in the $r-$range 
($\frac{1}{4}\eta \le r \le 1\eta$) explored here\footnote{A clarification 
is needed here to explain why the method of isolating caustic 
pairs successfully provides an $r-$independent kernel contribution 
for small equal-size particles with $St \lsim 1$. Like the case
of the weighted rms of caustic pairs, the mean relative velocity, 
$\langle w_{\rm r}\rangle^{\rm cau}$, per caustic pair is larger than the 
overall mean value. An important reason we can obtain an $r-$independent caustic kernel 
is that, as the continuous pairs are excluded, the number of caustic pairs is 
significantly smaller than the total number of pairs at $\frac{1}{4}\eta \lsim r\lsim 1\eta$. 
Thus, the contribution of caustic pairs, $g^{\rm cau}$, to the 
RDF is smaller than the overall RDF.  Fig.\ 18 of  Paper I shows that, 
as $r$ decreases, $\langle w_{\rm r}\rangle^{\rm cau}$ decreases,  
while the RDF $g^{\rm cau}$ increases.  Also, $g^{\rm cau}$ 
and $\langle w_{\rm r}\rangle^{\rm cau}$ decreases and increases with 
increasing threshold relative velocity, respectively. It was thus possible to select a 
threshold to make the caustic kernel contribution $g^{\rm cau} \langle |w_{\rm r}|\rangle^{\rm cau}$ 
$r-$independent.}. 
In this range of $r$, the majority of particle pairs with 
$St \lsim 0.4$ belong to the continuous type, and the accuracy of approximating 
the actual statistics in the $r\to 0$ limit by the collision velocity of caustic pairs is poor. 
As $r$ decreases to a sufficiently small value so that most particle pairs 
are of the caustic type, the problem is expected to be less severe.

The above problem concerning the use of caustic pairs to 
approximate the $r\to0$ statistics adds further uncertainty to the 
validity and accuracy of the criterion we adopted to isolate the caustic pairs, 
in addition to those already discussed in \S 3.3. Nevertheless, the classification of two types of particle pairs and 
the method used to splitting them  based on theoretical models do 
offer physical insight to understand the $r-$dependence of the kernel in the $r$ 
range we explored. It is still possible that, despite this problem with the weighted rms collision velocity, our 
method to obtain the $r-$independent caustic kernel provides 
a reasonable approximation for the kernel at $r\to0$ (see \S 3.3). 
Larger simulations that can measure the particle statistics at small scales are needed to 
test this possibility  and to resolve the convergence issue. 

We also note the red lines in the left panel of Fig.\ \ref{collisionv}  rise 
slightly as $St$ decreases toward $St =0.1$. This behavior is unexpected, and is likely a 
numerical artifact, as the trajectory integration of  $St=0.1$ 
particles, the smallest in our simulation, is the least accurate (see Papers I and II).  
The integration accuracy depends on the ratio of the simulation time step 
to the particle friction time, which is the largest for the smallest particles.  
This issue can be solved by future simulations with a better temporal resolution 
(see Paper II).   

\subsection{The Collision-Rate Weighted PDF}

In Paper III,  we computed  the unweighted PDF of turbulence-induced 
relative velocity for all Stokes number pairs available in our simulation, 
and showed that the PDF of ${\bs w}$ is generically non-Gaussian, 
exhibiting fat tails. In that paper, we discussed in details the trend of 
non-Gaussianity with the particle friction times, and interpreted the 
results using the physical picture of the PP10 model. Here we examine the collision-rate 
weighted PDF of $|{\bs w}|$. 

\begin{figure}[h]
\center{\includegraphics[height=2.9in]{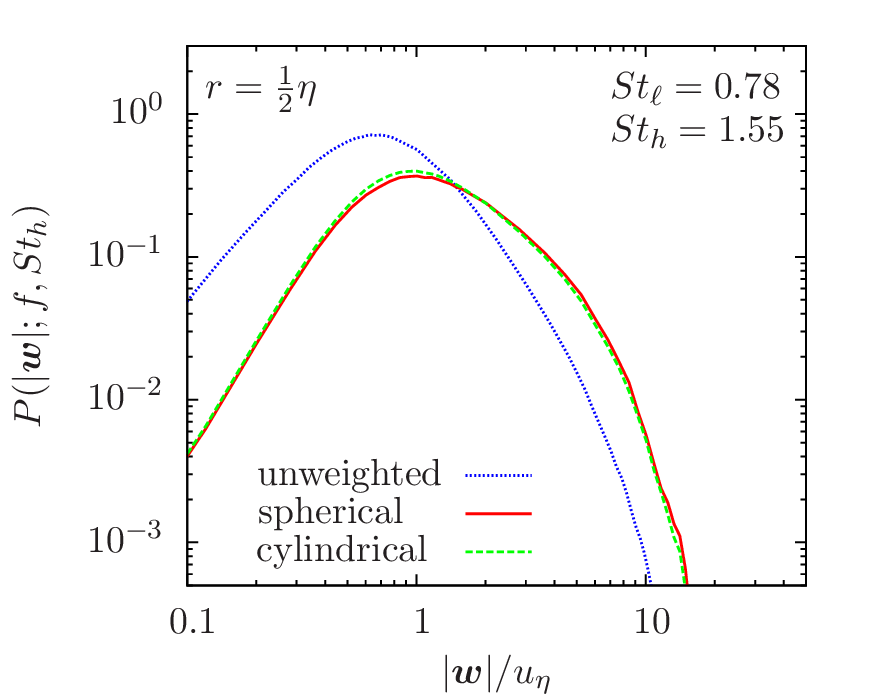}}
\caption{PDFs of the 3D relative velocity amplitude for particles with $St_\ell = 0.78$ 
and $St_h=1.55$ at a distance $r=\frac{1}{2} \eta$.  The blue line shows the unweighted PDF, 
while the red and green lines correspond  to the collision-rate weighted PDFs in the spherical and cylindrical 
formulations, respectively. 
}
\label{weightedPDF} 
\end{figure}

In Fig.\ \ref{weightedPDF}, we illustrate the effect of collision-rate weighing 
using a Stokes number pair, $St_{\ell} =0.78$ and $St_{h} =1.55$, as an example. 
The dotted line shows the unweighted PDF, while the solid and dashed lines are 
the weighted PDFs, $P_{\rm sph}$ and $P_{\rm cyl}$, in the spherical and cylindrical 
formulations, respectively. Clearly, with the weighting factor, 
the PDF shows lower and higher probabilities at small and large collision speeds, 
respectively. As it favors large-velocity collisions, the collision-rate weighting 
increases the fraction of collisions leading to fragmentation, and reduces the probability of sticking. 
Consequently, the growth of dust particles would be more difficult 
than predicted  by dust coagulation models that ignore the 
collision-rate weighting. The solid line and dashed lines almost coincide with each other. 
For this Stokes pair, the direction of ${\bs w}$ is nearly isotropic, and the 
weighted PDFs in the spherical and cylindrical formations are expected to be equal.

\begin{figure*}[t]
\includegraphics[height=2.9in]{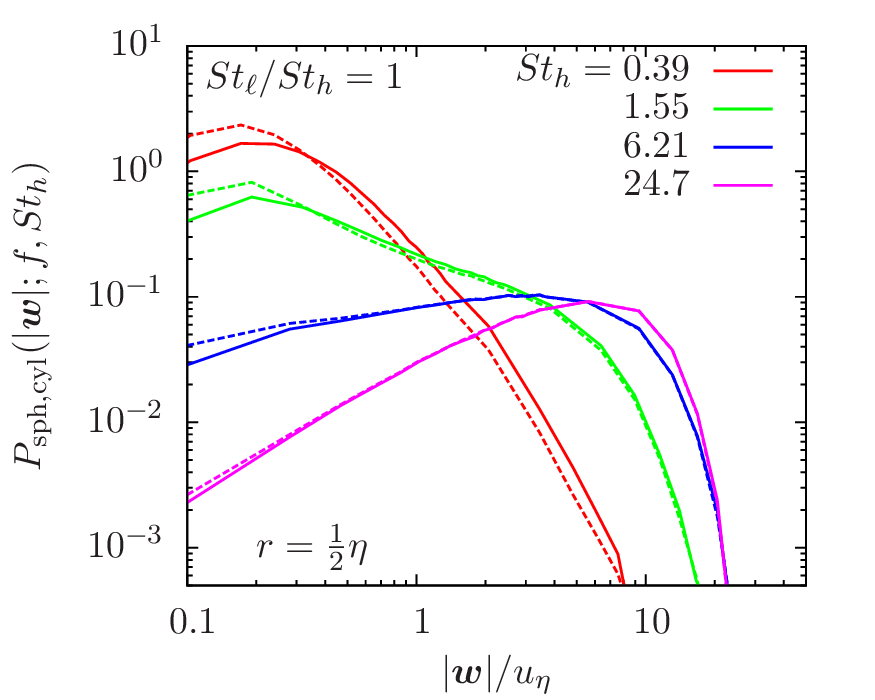}
\includegraphics[height=2.9in]{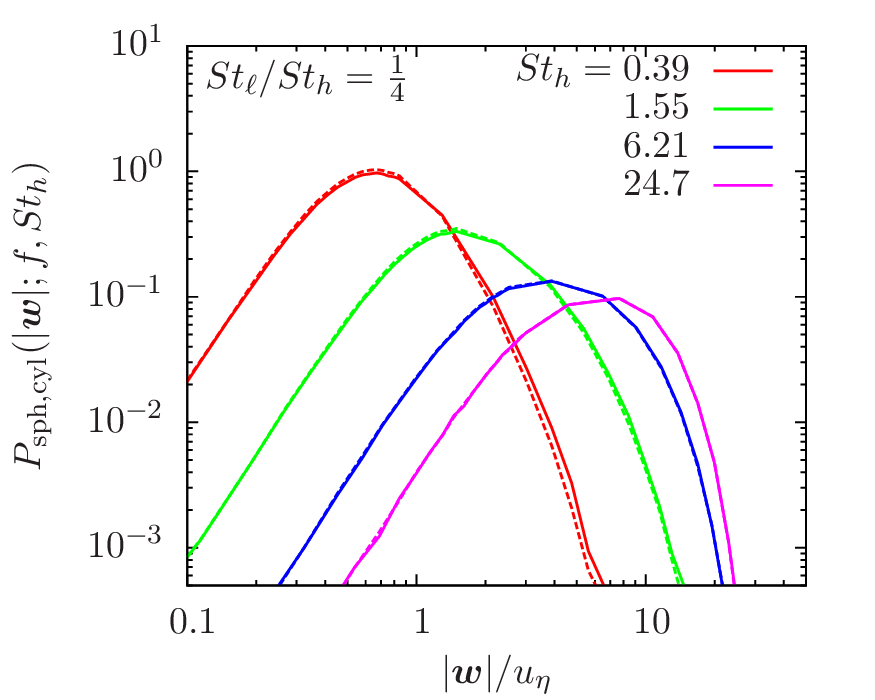}

\caption{Collision-rate weighted PDFs of the 3D amplitude, $|\bs{w}|$, in the spherical (solid) 
and cylindrical (dashed) formulations. The PDFs are measured at $r=\frac{1}{2} \eta$. 
The left panel plots the PDFs for particles of equal sizes, 
while the right panel shows results for a Stokes ratio of $f=\frac{1}{4}$.}
\label{comparepdf} 
\end{figure*}

Fig.\ \ref{comparepdf} plots $P_{\rm sph}$ (solid lines ) and $P_{\rm cyl}$ (dashed lines) at 
more Stokes number pairs. The left and right panels show the measured PDFs for $f=1$ 
and $\frac{1}{4}$, respectively. The weighed PDFs in the two formulations are in good 
agreement for all the Stokes pairs. In cases where the two PDFs show differences, $P_{\rm cyl}$ has 
slightly higher and lower probabilities than $P_{\rm sph}$ at small and large $|{\bs w}|$, 
respectively. For $f=1$, the difference between $P_{\rm sph}$ and $P_{\rm cyl}$ is largest 
at $St = 0.79$, and then decreases with increasing $St$. 
A comparison of the two panels shows that $P_{\rm sph}$ and $P_{\rm cyl}$ are in 
better agreement at smaller $f$.  As shown in Paper II, the isotropy of ${\bs w}$ with 
respect to ${\bs r}$ improves with increasing $St$ or decreasing $f$.

The finding that  $P_{\rm sph} \simeq P_{\rm cyl}$ at all 
Stokes number pairs suggests that  either formulation can be used to 
determine the collision outcome in coagulation models 
for dust particle growth. The geometry assumed in the spherical formulation is physically more reasonable, 
but the weighted PDF in the cylindrical formulation appears to be more convenient, 
as it is directly related to the unweighted PDF as $|{\bs w}| P(|{\bs w}|)/\langle|{\bs w}| \rangle$. 
With this relation, one can easily determine the 
shape of $P_{\rm cyl}(|{\bs w}|)$ from the unweighted PDF presented 
in Paper III. It is expected that the trend of the shape 
of $P_{\rm cyl}(|{\bs w}|)$ (and hence $P_{\rm sph}$) with $f$ and $St_{h}$ would be similar 
to the unweighted PDFs discussed in Paper III.  For example, if the shape of the unweighted PDF of $|{\bs w}|$ 
has a higher degree of non-Gaussianity, the weighted PDF would 
also have more probabilities at both left and right tails, corresponding 
to small and large collision velocities. We refer the interested reader to Paper III for a 
detailed discussion on the shape and the non-Gaussianity of the unweighted PDF 
as a function of $f$ and $St_h$.   

An important application of the collision velocity PDF is to 
determine the outcome of dust particle collisions induced by turbulence in protoplanetary disks. 
In Paper III, we computed the fractions of collisions leading to sticking ($F_{\rm s}$), bouncing 
($F_{\rm b}$) and fragmentation ($F_{\rm f}$) using typical disk and turbulence parameters at 1AU, and 
particular attention was given to the effect of the non-Gaussianity of the relative velocity PDF on the fractions.  
These fractions need to be incorporated in coagulation models for a realistic prediction 
of the dust size evolution. The calculation of the fractions was based on the PDF of $|{\bs w}|$ 
measured in our simulation using a weighting factor ($\propto |{\bs w}|$) from the cylindrical formulation. 
Several simplifying assumptions were used to extrapolate the measured PDFs in the 
simulated flow to the disk turbulence with much larger Reynolds number (see Paper III). 
In Appendix A, we compute $F_{\rm s}$, $F_{\rm b}$ and $F_{\rm f}$ from the weighted 
PDF, $P_{\rm sph} (|{\bs w}|)$, in the spherical formulation and compare the result with the cylindrical formulation. 
The calculation in Appendix A adopts the exact same parameters and assumptions as in Paper III.  
In Fig.\ \ref{fractions} of Appendix A, we see that $F_{\rm s}$, $F_{\rm b}$ and $F_{\rm f}$ computed 
from the two formulations are consistent with each other, 
as expected from the agreement between $P_{\rm sph}$ and $P_{\rm cyl}$.  
 
The spherical and cylindrical formulations yield very similar results for all the 
statistical quantities discussed so far, including the collision kernel, the 
collision-rate weighted rms and PDF of the 3D amplitude, $|\bs w|$, of the relative velocity, 
and the fractions of collisions leading to sticking, bouncing and fragmentation (Appendix A). 
All these quantities computed from the two formulations coincide 
with each other, except for small differences for Stokes pairs with $f$ close to 1 and $St_h \lsim 1$.

In Appendix B, we find an interesting difference between the two formulations concerning 
the collision-rate weighted PDFs of the radial and tangential components of ${\bs w}$. 
The statistics of the radial and tangential components is of interest because the 
collision outcome may depend on the angle at which the two particles collide. 
Appendix B  shows that, in the spherical formulation, the weighted PDF, $P_{\rm sph} (|w_{\rm r}|)$, of 
the radial velocity amplitude of approaching particle pairs is very close to the PDF, $P_{\rm sph} (w_{\rm tt})$, 
of the {\it total} amplitude, $w_{\rm tt}$, of the two tangential components. 
This suggests that, on average, the collision energy in the radial direction 
is the same as the total tangential collision energy. On the other hand, the 
weighted radial PDF,  $P_{\rm cyl} (|w_{\rm r}|)$, in the cylindrical formulation 
is found to almost coincide with the PDF, $P_{\rm cyl} (|w_{\rm t}|)$, of the amplitude, $|w_{\rm t}|$, 
of each {\it single} tangential component\footnote{As shown in Appendix B, the equalities $P_{\rm sph} (|w_{\rm r}|) = P_{\rm sph} (w_{\rm tt})$ 
and $P_{\rm cyl} (|w_{\rm r}|) = P_{\rm cyl} (|w_{\rm t}|)$ are expected if the direction of ${\bs w}$ is isotropically distributed.}. 
Thus, each of the three directions provides equal 
amount of collision energy, which is in sharp contrast to the spherical formulation. 
As the weighting factor $\propto w_{\rm r}$, the spherical formulation 
gives  more weight to collisions with larger radial relative speed. 
Consequently, it favors head-on collisions, and  predicts more 
head-on collisions than the cylindrical formulation.  

Different predictions for the weighted radial and tangential PDFs by the two 
formulations can be used to test which one is physically more realistic. 
The test requires directly measuring the PDFs of the radial and 
tangential relative speeds for particle pairs that are actually colliding. 
The direct measurement can be conducted using the method of Wang et al.\ (2000), 
which counts how many pairs of particles of given finite sizes get into contact 
during each small time step. Analyzing the statistics of the radial and 
tangential collision velocities of particle pairs in contact and comparing 
the results with the predictions by the two formulations would 
tell which formulation provides a better prescription 
for turbulence-induced collisions. 

\section{Discussion}

A major result of our study is that, due to their stronger clustering, 
the collision kernel for particles of similar sizes in the inertial range 
is larger than between different particles. This is not captured by 
the collision formulation commonly used in dust coagulation models, 
as the clustering effect is neglected. We discuss the implication 
of this result on dust particle growth in protoplanetary disks. 
As an illustrating example, we consider collisions of silicate particles at 1 AU in 
a minimum mass solar nebula. In the example, we use the 
same disk and turbulence parameters as in Paper III, to which 
we refer the reader for details. 

As the particle size grows to $\simeq 1$ mm at 1AU, the typical collision 
velocity induced by turbulence exceeds the bouncing threshold ($\simeq 5$ cm s$^{-1}$), 
and the collision outcome starts to be dominated by bouncing rather than 
sticking (Fig.\ \ref{fractions} in Appendix B; see also Paper III). 
If the probability distribution of the collision velocity is neglected, the 
growth of dust particles would stall at the millimeter size, a problem known as the bouncing 
barrier for planetesimal formation (Zsom et al.\ 2010, G\"ultler et al.\ 2010). The friction time, $\tau_{\rm p}$, 
of millimeter-size particles is close to the Kolmogorov time at 1 AU, 
meaning that the bouncing barrier starts once the particles reach the 
inertial range of the flow.    

Recent studies by Windmark et al.\ (2012) and Garaud et al.\ (2013) showed that the 
bouncing barrier may be overcome if the probability distribution of the collision velocity 
is taken into account. With a collision velocity distribution, there is always a finite 
probability of low-velocity collisions leading to sticking, even if the typical 
collision velocity is already above the bouncing threshold. If, by ``luck", a particle 
experiences low-velocity collisions continuously, it may grow significantly past 
the millimeter size.  
In Paper III, we showed that, unlike the assumption of Gaussian distribution 
made by Windmark et al.\ (2012) and Garaud et al.\ (2013), 
the PDF of the collision velocity ${\bs w}$ is highly non-Gaussian, and 
accounting for the non-Gaussianity gives significantly higher 
probabilities of sticking, further alleviating the bouncing barrier (see Paper III).  In 
Appendix B, we computed the probabilities 
of sticking, bouncing and fragmentation as a function of the particle size 
using both spherical and cylindrical formulations. 
   
A ``lucky" particle that grows past the bouncing barrier 
and reaches the fragmentation barrier around 
decimeter size can further grow by the so-called mass transfer mechanism 
(Windmark 2012). In this mechanism, a large particle acquires mass when 
colliding with much smaller particles. The collision breaks up the small particle, 
and some of its fragments stay on the large particle. The large particle can thus 
grow beyond the fragmentation barrier and toward the planetesimal size by 
continuously sweeping up mass of small particles (Windmark 2012). 
The mass transfer mechanism occurs only when the size ratio of the two particles 
is sufficiently large, and thus relies on the formation of large ``seed" by 
coagulational growth in between the bouncing and fragmentation barriers. 

Windmark et al.\ (2012) found that, due to the small 
sticking probability, particle growth from the bouncing barrier size  
toward the fragmentation barrier is very slow with a timescale 
of $10^4$ yr. In the present work, we have shown that turbulent 
clustering increases the particle collision rate, and thus can 
accelerate particle growth in between the two barriers, further alleviating 
the bouncing barrier. The acceleration is the most efficient for particles 
of similar sizes, for which the clustering effect peaks and the collision 
kernel is the largest. Interestingly, for collisions between particles 
of similar sizes, the probability of sticking is the largest. 
As shown in Fig.\ \ref{fractions} in Appendix B (see also Fig.\ 16 in Paper III), 
the decay of the sticking probability with increasing size, $a_{\rm p,h}$, 
of the larger particle is the slowest at $f=1$, and is significantly more 
rapid as $f$ decreases toward 0. These suggest that the effect of 
clustering preferentially increases the rate of collisions that 
have a higher probability of sticking. In other words, 
the clustering effect increases not only the overall collision rate, 
but also the overall fraction of collisions leading to sticking. 
Therefore, the acceleration of the particle growth by the clustering effect 
past the bouncing barrier is likely quite efficient. 
 The above discussion also indicates that particle growth between 
the bouncing and fragmentation barriers would occur mainly 
through collisions between particles of similar sizes.

The effect of turbulent clustering on dust particle growth 
at 1 AU can be summarized as follows. When 
reaching millimeter size, the particle friction time enters the 
inertial range of the flow, and turbulent clustering 
starts to significantly increase the collision kernel.  
Meanwhile, the collision outcome is dominated by bouncing, and 
the particle growth beyond the bouncing barrier relies on the finite 
but decaying probabilities of low collision velocity that allows sticking. 
Turbulent clustering helps accelerate the particle growth past the bouncing barrier, 
and, in particular, it preferentially enhances the collision rate between 
similar-size particles, which have a higher sticking probability. 
Accounting for the effect of clustering would thus help alleviate the 
bouncing barrier and accelerate the formation of large seed particles that 
can further grow beyond the fragmentation barrier by the sweep up mechanism. 

A quantitative estimate for how fast  turbulent clustering can 
accelerate the particle growth requires future studies to examine 
the Reynolds number ($Re$) dependence, which is needed to extrapolate the 
collision statistics in our flow to protoplanetary turbulence.  
Fig.\ \ref{rdf} shows that clustering effect is strongest  at  
$St \simeq 1$ (corresponding to $\simeq 1$mm at 1AU), and it can amplify the collision rate of 
equal-size particles with $St \simeq 1$ by an order of magnitude. 
The acceleration would be faster if the radial distribution function, $g$, has a significant $Re-$dependence. 
Note that in our simulation the clustering effect significantly decreases as $St$ increases above 1 and already becomes weak for $St =6.21$ 
particles in the inertial range (see Fig.\ \ref{rdf}). How much turbulent clustering can accelerate the 
collision rate of inertial-range particles of similar sizes in a realistic disk flow depends on the Reynolds number dependence of the RDF (see \S 3.5).    
A refined dust coagulation model that incorporates the clustering effect as 
well as the non-Gaussianity of the collision velocity is needed to give a conclusive 
answer for how and how fast dust particles grow in between the bouncing and 
fragmentation barriers. 
We point out that turbulent clustering may not be helpful for the fast 
drift problem of meter-size particles. The friction timescale 
of these large particles is close to the large eddy time in the disk, and the 
effect of turbulent clustering is expected to be weak, 
and would thus not help speed up the collisions of these particles.

\section{Summary and Conclusions}

Motivated by the problem of dust particle growth in protoplanetary disks, 
we studied the collision statistics of inertial particles suspended 
in turbulent flows. Using a numerical simulation, we evaluated the 
collision kernel as a function of the friction times of two particles 
of arbitrary sizes, accounting for the effects of turbulent clustering and 
turbulence-induced collision velocity. We also computed the statistics 
of the collision velocity using a collision-rate weighting factor to account for 
the higher collision frequency for particle pairs with 
larger relative speed. Below we list the main conclusions of this study.   

\begin{enumerate}

\item{If the friction time, $\tau_{\rm p,h}$, of the larger particle is below the Kolmogorov time, 
$\tau_\eta$, or above the Lagrangian correlation time, $T_{\rm L}$, the collision kernel 
per unit cross section is found to be larger at smaller Stokes (or friction time) ratio $f$ ($\equiv St_\ell/St_h$). 
This is because, at given $\tau_{\rm p,h}$, the particle relative velocity increases with decreasing $f$, 
due to the increase of a contribution, named the acceleration contribution, corresponding to 
different responses of particles of different sizes to the flow velocity. 
On the other hand, for $\tau_{\rm p,h}$ in the inertial range 
($\tau_\eta \lsim \tau_{\rm p,h} \lsim T_{\rm L}$), the kernel per unit cross section increases with increasing $f$ and 
is the largest at  $f=1$, due to the effect of turbulent clustering, 
which peaks for equal-size particles. Thus, as the typical particle friction time, 
$\tau_{\rm p}$, enters the inertial range of the flow, collisions between similar particles become more frequent 
than between particles of different  sizes. 
If the friction time, $\tau_{\rm p,h}$, of the larger particle  is close to $T_{\rm L}$, the kernel per unit cross section 
is about equal to the 1D rms flow velocity, $u'$, independent of $f $ or the friction time, $\tau_{\rm p,l}$, of the smaller particle.}

\item{The kernel formula commonly used in dust coagulation models neglects the effect of turbulent clustering and uses 
the rms relative velocity ($\langle w^2\rangle^{1/2}$) instead of the mean relative speeds ($\langle |w_{\rm r}|\rangle$ or $\langle |{\bs w}|\rangle$). 
The former underestimates the collision rate, while the latter tends to overestimate 
it, as the rms relative velocity  is larger than the mean. 
For particles of similar sizes with $St_{\rm h}$ in the inertial range, clustering is strong and more than compensates the latter effect. 
Neglecting clustering would thus significantly underestimate the collision rate of particles of similar sizes (${\bs f\simeq 1}$), especially 
at  $St_h \simeq 1$.}

\item{We analyzed the collision-rate weighted statistics for the 
particle relative velocity, ${\bs w}$. With a collision-rate weighting factor, 
the rms relative velocity measures the average collision energy per collision. 
The weighting factor is proportional to the particle relative velocity ($w_{\rm r}$ or $|{\bs w}|$), 
and favors collisions with larger collision velocity, leading to 
larger rms values than without weighting. The weighted rms corresponds to 
the 3rd order statistics, and the enhancement over the unweighted rms 
depends on the PDF shape of ${\bs w}$. 
The ratio of the weighted rms to the unweighted one increases as $f$ increases toward 1, 
due to the fatter PDF shape at larger $f$. The ratio is the largest for particles of equal sizes ($f=1$), 
and ignoring the collision-rate weighting would significantly underestimate the average 
collision energy between particles of similar sizes. For $\tau_{\rm p,h}$ in the inertial range, 
the weighted rms is  independent of the size of the smaller particle, and scales 
with $\tau_{\rm p,h}$ roughly  as $\tau_{\rm p,h}^{1/2}$ for any $f$. For particles with 
$\tau_{\rm p,h} \simeq T_{\rm L}$ in our simulation, the weighted rms collision velocity 
is approximately $\simeq 1.3 u'$ at any $f$. 

As more weight is given to collisions with larger relative velocity, the weighted distribution of the 
3D amplitude, $|{\bs w}|$, of the collision velocity shows higher (lower) probabilities at large (small) 
collision velocities. Therefore, ignoring the collision-rate weighting would overestimate (or underestimate) 
the fraction of collisions resulting in sticking (or fragmentation).} 

\item{We argued that the effect of turbulent clustering helps alleviate the bouncing barrier for 
planetesimal formation. As it increases the probability of finding colliding neighbors, 
the spatial clustering of dust particles enhances the collision kernel, and can 
accelerate the particle growth past the bouncing barrier. The acceleration is the 
fastest for particles of similar sizes in the inertial range, and, considering that the probability of 
sticking is also the largest for particles of similar sizes, the effect of 
clustering not only increases the overall collision rate, but also the overall fraction of 
collisions leading to sticking. The acceleration of particle growth by turbulent clustering beyond the bouncing 
barrier is thus expected to be efficient.}

\item{We compared two kernel formulations based on spherical and cylindrical 
geometries. The two formulations give consistent results 
for the collision rate, the collision-rate weighted rms and PDF of $|{\bs w}|$. 
We showed that these quantities computed from the two formulations
are expected to be exactly equal, if the direction of the relative velocity, $\bs w$, 
of two particles is statistically isotropic with respect to their separation, ${\bs r}$. 
For small particles of similar sizes, ${\bs w}$ is not perfectly isotropic, 
and slight differences between the two formulations are observed  at $f \gsim \frac{1}{2}$ and $St_{h} \lsim 1$. 
Isotropy of $\bs w$ improves with increasing $St_h$ or decreasing $f$, 
leading to the equality of the collision kernel and collision-rate 
weighted PDFs of $|\bs w|$ in the two formulations. 
The two formulations also give similar estimates 
for the fractions of collisions resulting in sticking, bouncing and fragmentation.

An interesting difference between the two formulations is found concerning 
the weighted PDFs of the radial and tangential components of ${\bs w}$. 
In the spherical formulation, the weighted PDF of the radial component 
almost coincides with that of the {\it total} tangential amplitude, while in the cylindrical formulation 
the weighted radial PDF is about equal to the distribution of each {\it single} tangential component.  
As its weighting factor favors collisions with larger radial collision speed, the spherical 
formulation predicts more head-on collisions than the cylindrical formulation. 
This difference provides a unique test for the two kernel prescriptions with different geometries.    
}
\end{enumerate}

Although the present study draws significant conclusions on the statistics 
of dust particle collisions induced by turbulence, several important questions 
remain to be answered by future work.  First, the Reynolds-number dependence 
of the clustering effect and the collision kernel is currently unknown and needs to be 
examined using simulations at higher resolutions. Application of the 
measured statistics in our simulation to protoplanetary disks requires an 
extrapolation to high Reynolds numbers appropriate for the disk conditions. 

Second, our simulation data provides accurate statistical 
measurements only at particle distances $r\gsim \frac{1}{4}$, 
but the collision kernel for small particles of equal-size with $St \lsim 1$ 
does not converge at $r=\frac{1}{4}\eta$. We have discussed some uncertainties 
in our method to isolate an $r-$ independent 
kernel contribution by separating out the caustic particle pairs. 
Also, the collision-rate weighted rms relative velocity for small particles of the same size has not reached 
convergence at $r\simeq \frac{1}{4}\eta$. Simulations with a larger number 
of particles that allow measurements at smaller $r$ would help resolve these issues.   

Another interesting question is whether and how the 
collision statistics of particles with friction time that couples 
to the flow driving scales is affected by the specific driving mechanism. 
A set of numerical simulations with different driving schemes would help clarify 
the possible effects of the driving force.

Finally, to evaluate the accuracy of the collision rate commonly 
used in the dust coagulation models, we need to carefully test the 
prediction of the model by V\"olk et al.\ (1980) and its successors for the (unweighted) 
rms relative velocity. We will conduct a systematic test of the V\"olk-type 
models against our simulation data in a future paper. 

\acknowledgements 

Resources supporting this work were provided by the NASA High-End Computing (HEC) Program through the NASA Advanced 
Supercomputing (NAS) Division at Ames Research Center, and by the Port d'Informaci— Cient'fica (PIC), Spain, maintained by a 
collaboration of the Institut de F'sica d'Altes Energies (IFAE) and the Centro de Investigaciones EnergŽticas, Medioambientales y
Tecnol—gicas (CIEMAT). LP is supported by a Clay Fellowship at Harvard-Smithsonian Center for Astrophysics. 
PP acknowledges support by the FP7-PEOPLE- 2010-RG grant PIRG07-GA-2010-261359.

\appendix

\section{Comparing Collision Outcomes in Spherical and Cylindrical Formulations}

In Paper III, we computed the fractions, $F_{\rm s}$, $F_{\rm b}$, 
and $F_{\rm f}$, of collisions leading to sticking, bouncing and 
fragmentation as a function of the dust particle sizes in protoplanetary disks, 
using the relative velocity PDF measured in our simulation. 
In the calculation, we adopted a minimum mass solar nebula and considered particle 
collisions at 1AU as an illustrating example. Turbulence condition in the disk was 
specified using the prescription of Cuzzi et al.\ (2001) with the Shakura-Sunyaev 
parameter $\alpha_{\rm t}$ set to $10^{-4}$. Due to the limited resolution, 
the inertial range of our simulated flow is very short, and we made a few assumptions 
to extrapolate the measured PDFs to the disk flow with a much larger Reynolds number. 
For example, 
we set the shape of the PDF for all inertial-range particles 
in a real disk to be same as the measured distribution of $St=6.21$ particles, 
the only particle species that definitely lies in the inertial range of the 
simulated flow. Clearly, this is a strong simplification, and
an accurate treatment needs the Reynolds-number 
dependence of the PDF, which is currently unknown. We assumed that, 
for collisions of silicate particles, bouncing and fragmentation occur as 
the collision velocity exceeds the threshold values of 5 cm s$^{-1}$ and 1 m s$^{-1}$, 
respectively. We refer the reader to Paper III for a detailed 
description of the assumptions and parameters used in the calculation. 

\begin{figure*}[h]
\centerline{\includegraphics[width=0.6\columnwidth]{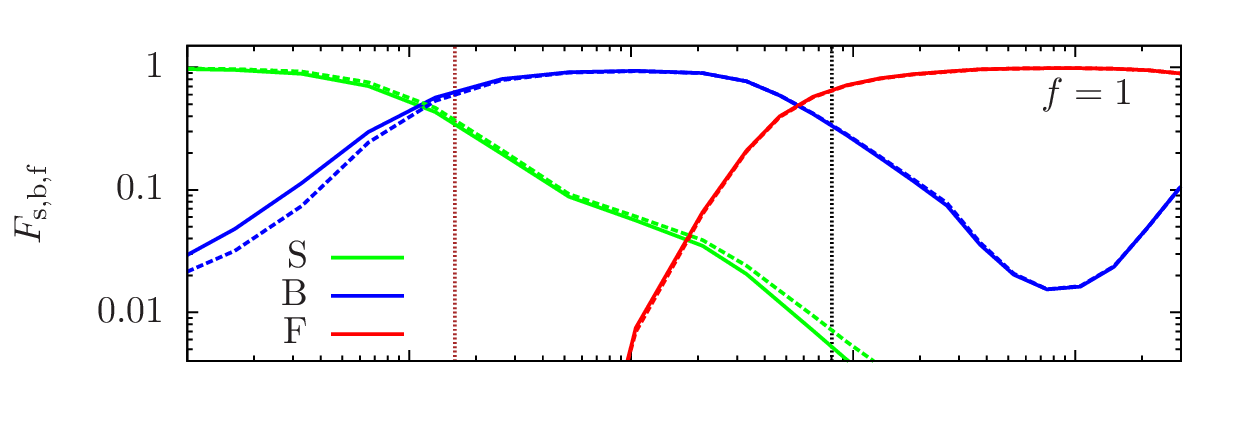}}
\vspace{-1.09cm}
\centerline{\includegraphics[width=0.6\columnwidth]{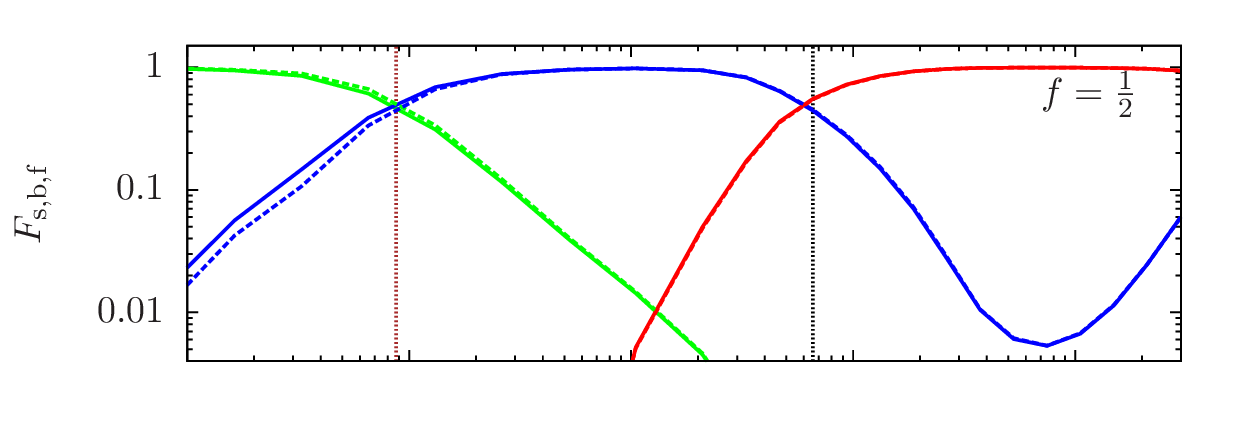}}
\vspace{-1.1cm}
\centerline{\includegraphics[width=0.6\columnwidth]{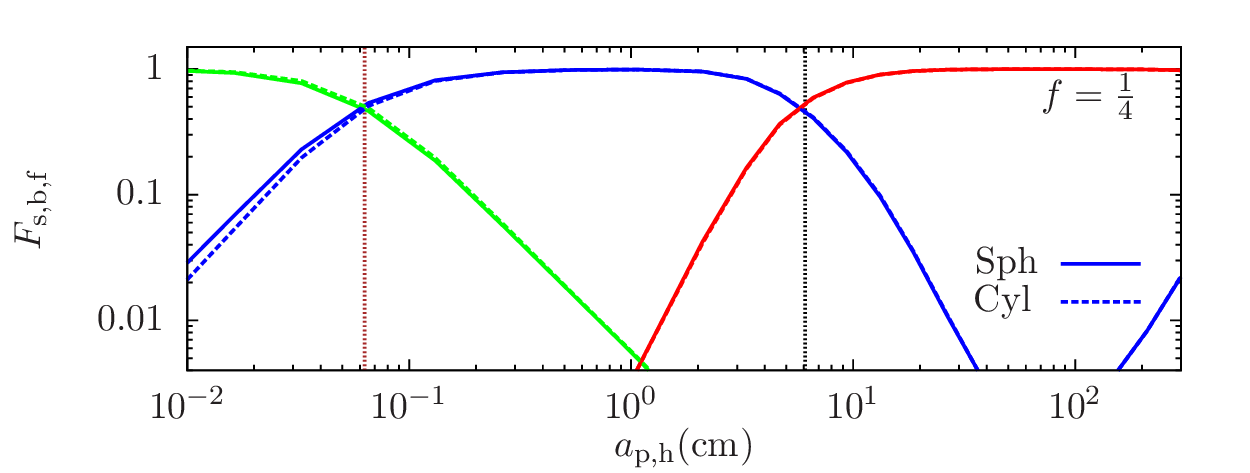}}
\caption{Sticking (green), bouncing (blue) and fragmentation (red) 
fractions as a function of the size, $a_{\rm {p, h}}$, of the larger particle. The three 
panels show fixed Stokes number ratios at $f=1$, $\frac{1}{2}$, and $\frac{1}{4}$.
 Vertical brown and black dotted lines correspond to critical $a_{\rm {p, h}}$ 
at which the unweighted rms relative velocity is equal to 
bouncing and fragmentation thresholds, respectively. Solid and dashed lines are 
computed from the weighted PDFs at a particle distance of $r=\frac{1}{2} \eta$ 
using the spherical and cylindrical formulations, respectively.  
}
\label{fractions} 
\end{figure*}

The calculation of $F_{\rm s}$, $F_{\rm b}$, and $F_{\rm f}$ in 
Paper III was based on the measured PDF of the 3D amplitude, $|{\bs w}|$, using a cylindrical weighting 
factor $\propto |{\bs w}|$. Here, we compare the 
estimates  from the spherical and cylindrical formulations. Fig.\ \ref{fractions} shows 
$F_{\rm s}$ (green), $F_{\rm b}$ (blue), and $F_{\rm f}$ (red), as a function of the size, $a_{\rm p, h}$, 
of the larger particle. The solid and dashed lines are calculated 
from the wighted PDFs, $P_{\rm sph}$ and $P_{\rm cyl}$, at a particle distance $r=\frac{1}{2}\eta$.  
As mentioned in Paper III, it is desirable to use PDFs at smaller $r$, 
which, however, cannot be measured at high accuracy 
in our simulation. The figure is plot in the same way as  Fig.\ 16 of 
Paper III. The three panels from top to bottom 
show the results for Stokes number ratios of $f=1$, $\frac{1}{2}$, and $\frac{1}{4}$, 
respectively.  The dashed lines here correspond to the solid lines 
in Fig.\ 16 of  Paper III.

The vertical brown and black dotted lines in each panel mark the sizes of the 
larger particle, at which the unweighted rms relative velocity reaches the 
bouncing and fragmentation thresholds, respectively. 
In dust coagulation models that ignore the collision velocity distribution, 
it was assumed that the collision outcome makes a sharp 
transition from sticking to bouncing and to fragmentation at the brown and dotted lines, 
respectively. Including the collision velocity distribution 
makes the transitions gradual because with a distribution 
there is always a finite probability of finding low (or high) collision velocities, even if 
the rms relative velocity is already above (or still below) the bouncing or fragmentation thresholds (see Windmark 2012, Garaud et al.\ 2013).  

In Paper III, we showed that the distribution of ${\bs w}$ is non-Gaussian, 
and accounting for the non-Gaussianity leads to more gradual transitions. 
In particular, the decay of the sticking fraction with increasing $a_{\rm p, h}$ 
is significantly slower than the estimate based on the assumption of a 
Gaussian distribution (see Fig.\ 16 of Paper III).  We therefore argued that
incorporating the non-Gaussianity 
of the collision velocity is helpful for the particle growth and further alleviates the problem of the bouncing barrier. 
A comparison of the three panels shows that the sticking fraction is 
more persistent at larger $f$. This corresponds to the higher degree of 
non-Gaussianity of the  PDF for particles of similar sizes.  
A detailed discussion on the behavior of the fractions with $f$ and $St_{\rm h}$ was given in Paper III.  

The solid and dashed lines for the spherical and cylindrical formulations almost coincide, 
consistent with the agreement of $P_{\rm sph}$ and $P_{\rm cyl}$ found in \S 4.2. 
This suggests that either formulation can be used to estimate the collision outcome. 
The slight difference between the solid and dashed lines  
corresponds to the difference between $P_{\rm sph}$ and $P_{\rm cyl}$ shown 
in Fig.\ \ref{comparepdf}. The difference decreases with 
decreasing $f$ because the direction of ${\bs w}$ is more isotropic at smaller $f$ and 
the agreement between $P_{\rm sph}$ and $P_{\rm cyl}$ improves (see Fig.\ \ref{comparepdf}).

\section{Collision-rate Weighted PDFs in Radial and Tangential Directions}

In this appendix, we consider the radial and tangential 
components of the relative velocity. An analysis of the radial and tangential relative speeds 
is of interest, as the collision outcome may depend on the angle at which two 
particles collide. We decompose ${\bs w}$ to a radial 
component, $w_{\rm r}$, and two tangential components, $w_{\rm t1}$ and $w_{\rm t2}$. 
In an isotropic flow, $w_{\rm t1}$ and $w_{\rm t2}$ are 
statistically equivalent, and we use $w_{\rm t}$ to represent the statistics of one tangential component. 
 In Papers I, II and III, we studied the statistics of $w_{\rm r}$ and $w_{\rm t}$ 
without collision-rate weighting. 
Here we examine the collision-rate weighted PDFs of $w_{\rm r}$ and $w_{\rm t}$. 
We also analyze  the total tangential amplitude, $w_{\rm tt}$, defined as $w_{\rm tt} \equiv (w_{\rm t1}^2 + w_{\rm t2}^2)^{1/2}$. 

In the spherical formulation, the weighted distribution, $P_{\rm sph} (|w_{\rm r}|)$, for the 
radial amplitude, $|w_{\rm r}|$, of approaching particle pairs ($w_{\rm r} \le 0$) is 
related to the left (negative) wing of the unweighted radial PDF, $P(w_{\rm r})$, 
by $P_{\rm sph} (|w_{\rm r}|)\propto |w_{\rm r}| P(-|w_{\rm r}|)$. In terms of the PDF, $P_{\rm v}({\bs w})$, 
of the vector ${\bs w}$, we have,  
\begin{gather}
P_{\rm sph} (|w_{\rm r}|) = -\frac{1}{N}\int\limits_{-\infty}^0  w'_{\rm r} dw'_{\rm r} \int\limits_{-\infty}^{\infty} dw'_{\rm t1} 
\int\limits_{-\infty}^{\infty} dw'_{\rm t2}  \delta(|w_{\rm r}|+w'_{\rm r} ) P_{\rm v} (w'_{\rm r}, w'_{\rm t1}, w'_{\rm t2}),
\label{sphradialpdf}
\end{gather}
where $N=- \int_{-\infty}^{0} w_{\rm r} P(w_{\rm r}) dw_{\rm r}$ normalizes the total probability to unity.  
The weighted PDF for the total tangential amplitude, $w_{\rm tt}$, is given by, 
\begin{gather}
P_{\rm sph} (w_{\rm tt}) = -\frac{1}{N}\int\limits_{-\infty}^0 w'_{\rm r} dw'_{\rm r} \int\limits_{-\infty}^{\infty} dw'_{\rm t1} 
\int\limits_{-\infty}^{\infty} dw'_{\rm t2}   \delta(w_{\rm tt}- (w^{\prime 2}_{\rm t1}+w^{\prime 2}_{\rm t2})^{1/2} ) 
P_{\rm v} (w'_{\rm r}, w'_{\rm t1}, w'_{\rm t2}).
\label{sphtanpdf}
\end{gather}
Again, it is useful to consider the case where ${\bs w}$ is isotropic and 
$P_{\rm v} ( w_{\rm r},  w_{\rm t1},  w_{\rm t2}) = P_{\rm v} (|{\bs w}|) = P_{\rm v} ((w_{\rm r}^2 + w_{\rm t1}^2 + w_{\rm t2}^2)^{1/2})$.  
In this case, integration of eqs.\ (\ref{sphradialpdf}) and (\ref{sphtanpdf}) gives, 
\begin{gather}
P_{\rm sph} (|w_{\rm r}|) = \frac{2\pi}{N} |w_{\rm r}|\int_{|w_{\rm r}|}^{\infty} |{\bs w}| P_{\rm v} ( |{\bs w}|)d  |{\bs w}|, \hspace{0.2cm} {\rm and}\hspace{0.2cm} 
P_{\rm sph} (w_{\rm tt}) = \frac{2\pi}{N} w_{\rm tt}\int_{w_{\rm tt}}^{\infty} |{\bs w}| P_{\rm v} ( |{\bs w}|)d  |{\bs w}|. 
\label{isopdfradtan}
\end{gather} 
Therefore, if  the direction of ${\bs w}$ is uniformly distributed with respect to ${\bs r}$, the collision-rate 
weighted PDF of the radial component is identical to that of the total tangential amplitude. 

\begin{figure*}[t]
{\includegraphics[height=2.95in]{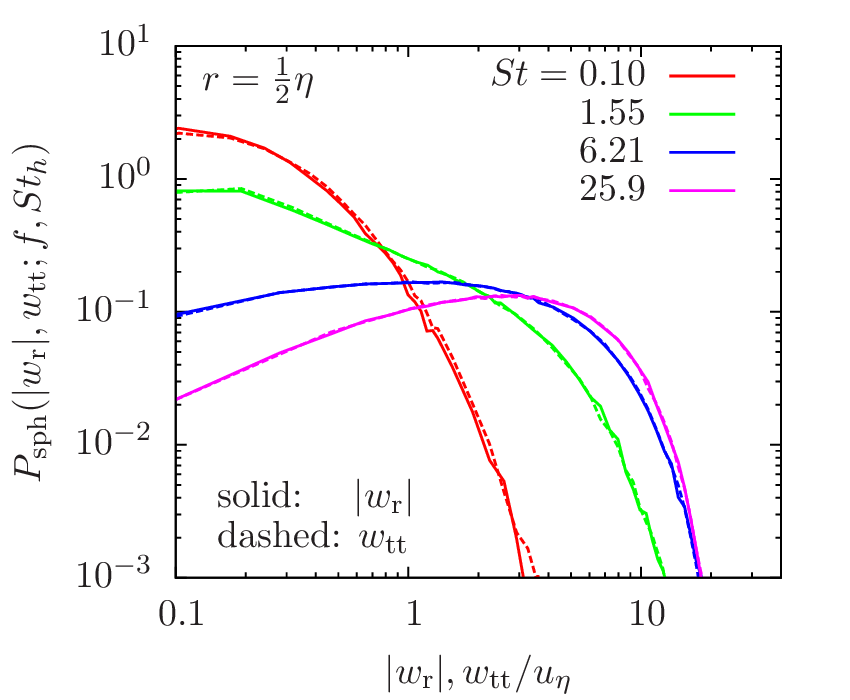}}
{\includegraphics[height=2.95in]{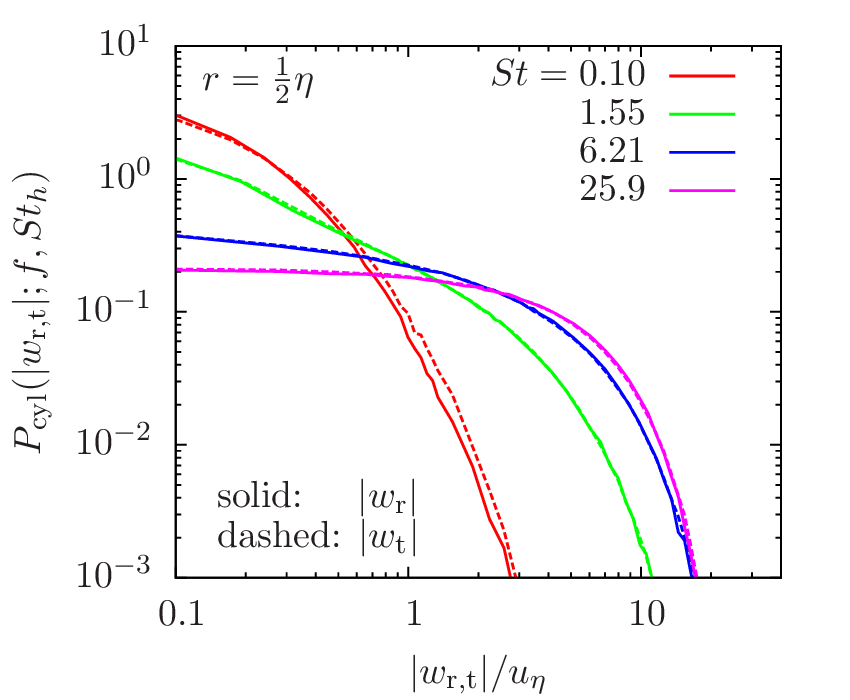}}
\caption{Left panel: Collision-rate weighted PDFs for the amplitude, $|w_{\rm r}|$, of the radial relative 
speed and the {\it total} tangential amplitude, $w_{\rm tt}$ for equal-size particles with different $St$. 
Right panel: the weighted PDFs of the amplitudes of the radial component, $|w_{\rm r}|$, and of a {\it single} 
tangential component in the cylindrical formulations. In both panels, the PDFs are measured at $r=\frac{1}{2} \eta$. 
}
\label{PDFrt} 
\end{figure*}

In the left panel of Fig.\ \ref{PDFrt},  we plot $P_{\rm sph} (|w_{\rm r}|)$ and $P_{\rm sph}(w_{\rm tt})$ 
in the spherical formulation for particles of equal sizes.  
The figure shows that the weighted PDFs of $|w_{\rm r}|$ and $w_{\rm tt}$ almost coincide for all $St$. 
Slight difference occurs only at small $St\lsim 1$, where the direction of ${\bs w}$ is not 
perfectly isotropic (see Paper I). As $St$ increases above 1, the direction of ${\bs w}$ 
becomes isotropic, and the difference between $P_{\rm sph} (|w_{\rm r}|)$ and 
$P_{\rm sph}(w_{\rm tt}$) disappears.  The isotropy of ${\bs w}$ also increases with 
decreasing Stokes ratio $f$ or decreasing particle distance, $r$ (Paper II). Therefore, 
a better agreement between $P_{\rm sph} (|w_{\rm r}|)$ and $P_{\rm sph}(w_{\rm tt})$ is 
observed at smaller $f$ and/or smaller $r$ (not shown).

The  near coincidence of $P_{\rm sph} (|w_{\rm r}|)$ and $P_{\rm sph}(w_{\rm tt})$ 
indicates that the weighted rms of the radial collision velocity is about equal to that of the 
total tangential amplitude, or equivalently, on average, the collision energy in the radial 
direction is the same as the total tangential energy. 
This is in contrast to the case without collision-rate weighting, 
where the rms of the radial relative speed t is equal to that of 
each tangential component, expect for slight differences for small particles of similar sizes 
with $St_h \lsim 1$ (see Papers I and II).   

As the collision rate is proportional to $|w_{\rm r}|$ in the spherical 
formulation, collisions with larger radial relative velocity are given 
more weight. The finding that the weighted rms and 
PDF of the radial component is the same as the total tangential amplitude  
suggests a higher frequency of head-on collisions than if the collision-rate weighting 
is ignored. In other words, the collision-rate weighting in the spherical 
formulation favors head-on collisions. This may have important implications 
for the collisional growth of dust particles if the collision outcome has a 
significant dependence on the colliding angle.  

We also computed the weighted PDFs of the radial and tangential 
components in the cylindrical formulation. Unlike the case of the spherical formulation, 
the weighting factor ($\propto|{\bs w}|$) in the cylindrical formulation does not 
favor a particular direction. Thus, if the direction of ${\bs w}$ is statistically isotropic, 
the weighted PDFs of the three components of the relative velocity are 
expected to the same. The weighted PDF of the radial amplitude, $|w_{\rm r}|$, in the 
cylindrical formulation can be defined  as $P_{\rm cyl} (|w_{\rm r}|) =  \frac{1}{\langle |{\bs w}| \rangle} \int_{-\infty}^{\infty}  dw'_{\rm r} \int_{-\infty}^{\infty} dw'_{\rm t1} 
\int_{-\infty}^{\infty} dw'_{\rm t2}  \delta(|w_{\rm r}| - |w'_{\rm r}|)  |{\bs w}|  P_{\rm v} (w'_{\rm r}, w'_{\rm t1}, w'_{\rm t2})$, 
and, assuming isotropy for the direction of ${\bs w}$,  
we find $P_{\rm cyl} (|w_{\rm r}|) = 4 \pi \int_{|w_{\rm r}|}^{\infty} |{\bs w}|^2P_{\rm v} (|{\bs w}|) d |{\bs w}|/\langle |{\bs w}| \rangle$.  
Similarly, it can be shown that the weighted PDF for the amplitude of {\it one} tangential component, $w_{\rm t}$, 
has the same form, $P_{\rm cyl} (|w_{\rm t}|) = 4 \pi \int_{|w_{\rm t}|}^{\infty} |{\bs w}|^2P_{\rm v} (|{\bs w}|) d |{\bs w}|/\langle |{\bs w}| \rangle$.
In the right panel of Fig.\ \ref{PDFrt}, we plot the weighted PDFs of $|w_{\rm r}|$ (solid lines) and 
$|w_{\rm t}|$ (dashed lines) for particles of equal size in the cylindrical formulation. 
As expected,  $P_{\rm cyl}(|w_{\rm r}|)$ and $P_{\rm cyl}(|w_{\rm t}|)$ 
coincide at $St \gsim 1$, where the direction of ${\bs w}$ is isotropic.  At $St \lsim 1$, 
the tail of $P_{\rm cyl}(|w_{\rm t}|)$ is slightly broader than $P_{\rm cyl}(|w_{\rm r}|)$, and, 
correspondingly, the weighted rms relative velocity in a tangential direction is slightly 
larger than in the radial direction. The equality of $P_{\rm cyl}(|w_{\rm r}|)$ and $P_{\rm cyl}(|w_{\rm t}|)$ 
also improves with decreasing $f$ and/or decreasing $r$.  
A comparison of the solid lines in the left and right panels of Fig.\ \ref{PDFrt} shows that  the two formulations 
give different estimates for the weighted radial PDFs.

The near equality of $P_{\rm cyl}(|w_{\rm r}|)$ and $P_{\rm cyl}(|w_{\rm t}|)$ in 
the cylindrical formulation suggests that, on average, each of three directions 
provides the same amount of collision energy.  
Thus, the total collision energy in the two tangential directions is twice larger than 
in the radial direction. This  is 
different from  the spherical formulation, where the collision 
energy in the radial direction is about equal to the 
sum from the two tangential components. The spherical formulation favors 
collisions with larger radial relative speed and predicts more head-on
collisions than the cylindrical formulation.   

In summary, we find an interesting difference between the two 
kernel formulations. In the spherical formulation, the weighted 
PDF of the radial component almost coincides with that of the 
{\it total} tangential amplitude, while in the cylindrical formulation the 
weighted radial PDF is nearly equal to the distribution of each 
{\it single} tangential component. Note that the two formulations give 
similar estimates for all the quantities discussed in the text and 
in Appendix A, including the collision kernel, the collision-rate weighted PDF of $|{\bs w}|$, 
and the fractions of collisions leading to sticking, bouncing and fragmentation. 
Thus, the difference in the predictions for the weighted 
radial and tangential PDFs provides a useful tool 
to test the two formulations. 
To carry out the test, one can directly 
measure the collision statistics by considering the finite particle size (instead of assuming point 
particles) and counting how many pairs of particles get into 
contact during each infinitesimal time interval.  Analyzing the radial and tangential 
relative velocities of these particle pairs in contact would tell which formulation 
is a better prescription for turbulence-induced collisions. 
 We defer the direct measurements of the collision statistics to a future work.

\end{document}